\documentclass[manuscript,nonacm]{acmart}
\usepackage{comment}
\usepackage{color}
\usepackage{algorithm}
\usepackage[noend]{algpseudocode}
\usepackage{amsmath}
\usepackage{balance}
\usepackage{caption}
\usepackage{subfig}
\usepackage{graphicx}
\usepackage{url}
\usepackage{hyperref}
\usepackage{caption}
\usepackage{pifont}
\usepackage{siunitx}
\usepackage[utf8]{inputenc}
\usepackage{hyperref}
\usepackage{enumitem}
\usepackage[printonlyused,withpage]{acronym}

\newcounter{TODO}

\newcommand{\nip}[1]{\vspace{1ex}\noindent\textbf{#1}}
\newcommand{\eg}{\emph{e.g., }}
\newcommand{\ie}{\emph{i.e., }}

\definecolor{light-gray}{gray}{0.95}
\newcommand{\code}[1]{\colorbox{light-gray}{\texttt{\small #1}}}
\newcommand{\spec}[1]{\textsl{#1}}

\newcommand{\hide}[1]{}

\usepackage{afterpage}
\usepackage{multicol}
\newsavebox{\shortpagebox}
\makeatletter
\newcommand{\shortpage}[1]
{\par
  \setbox\shortpagebox=\vbox{\strut #1\par}%
  \afterpage{\onecolumn
    \begin{multicols}{2}
    \unvbox\AP@partial
    \end{multicols}}%
  \unvbox\shortpagebox
\par}
\makeatother
\begin{document}
\title{Intel TDX Demystified: A Top-Down Approach}
\author{Pau-Chen Cheng}
\affiliation{
\institution{IBM Research}
\country{USA}
}
\email{pau@us.ibm.com}

\author{Wojciech Ozga}
\affiliation{
\institution{IBM Research}
\country{Switzerland}
}
\email{woz@zurich.ibm.com}

\author{Enriquillo Valdez}
\affiliation{
\institution{IBM Research}
\country{USA}
}
\email{rvaldez@us.ibm.com}

\author{Salman Ahmed}
\affiliation{
\institution{IBM Research}
\country{USA}
}
\email{sahmed@ibm.com}

\author{Zhongshu Gu}
\affiliation{
\institution{IBM Research}
\country{USA}
}
\email{zgu@us.ibm.com}

\author{Hani Jamjoom}
\affiliation{
\institution{IBM Research}
\country{USA}
}
\email{jamjoom@us.ibm.com}

\author{Hubertus Franke}
\affiliation{
\institution{IBM Research}
\country{USA}
}
\email{frankeh@us.ibm.com}

\author{James Bottomley}
\affiliation{
\institution{IBM Research}
\country{USA}
}
\email{jejb@us.ibm.com}

\renewcommand{\shortauthors}{IBM Research}
\begin{abstract}\normalsize
Intel \ac{TDX} is a new architectural extension in the 4th Generation Intel Xeon Scalable Processor that supports confidential computing. \ac{TDX} allows the deployment of virtual machines in the \ac{SEAM} with encrypted CPU state and memory, integrity protection, and remote attestation. \ac{TDX} aims to enforce hardware-assisted isolation for virtual machines and minimize the attack surface exposed to host platforms, which are considered to be untrustworthy or adversarial in the confidential computing's new threat model. \ac{TDX} can be leveraged by regulated industries or sensitive data holders to outsource their computations and data with end-to-end protection in public cloud infrastructure.

This paper aims to provide a comprehensive understanding of \ac{TDX} to potential adopters, domain experts, and security researchers looking to leverage the technology for their own purposes.
We adopt a top-down approach, starting with high-level security principles and moving to low-level technical details of \ac{TDX}. Our analysis is based on publicly available documentation and source code, offering insights from security researchers outside of Intel. 
\end{abstract}
\maketitle
\clearpage
\tableofcontents
\acresetall
\section{Introduction}
Deploying computations to cloud infrastructure can reduce costs, but regulated industries have concerns about moving sensitive data to third-party cloud service providers. Confidential computing aims to provide end-to-end protection for outsourced computations by minimizing the root of trust to the processors and their vendors. This means that all data must be protected throughout its life-cycle, from leaving its owners' premises to entering certified CPU packages in the cloud. Any adversaries, such as those intercepting on the network, disk storage, or main memory, should not be able to access the data in clear form. 

Cryptographic mechanisms, such as storage encryption and secure communication channels, protect the confidentiality, integrity, and authenticity of data both \emph{at rest} and \emph{in transit}. The emerging CPU-based \ac{TEE} techniques aim to provide protection for \emph{data in use}, \ie data loaded into main memory.

Intel \acf{TDX} is an architectural extension that provides \ac{TEE} capabilities in the 4th Generation Intel Xeon Scalable Processors. \ac{TDX} introduces the \ac{SEAM} to offer cryptographic isolation and protection for \acp{VM}, which are called \acp{TD} in the \ac{TDX} terminology. The threat model assumes that the privileged software, such as hypervisors or host operating systems, may be untrustworthy or adversarial. \ac{TDX} aims to protect the confidentiality and integrity of CPU state and memory for designated \acp{TD}, and also enables \ac{TD} owners to verify the authenticity of remote platforms. \ac{TDX} is built using a combination of techniques, including \ac{VT}~\cite{vtx}, \ac{MKTME}~\cite{mktmewhitepaper}, and the \ac{TDX} Module~\cite{tdxmodulespec}. \ac{TDX} also relies on \ac{SGX}~\cite{mckeen2013innovative} and \ac{DCAP}~\cite{dcapwhitepaper} for remote attestation. 

Throughout the paper, we aim to give an objective review of \ac{TDX}. Our goal is to provide a thorough understanding of \ac{TDX} to potential adopters, domain experts, and security researchers who want to leverage or investigate the technology for their own purposes. All the information is based on publicly available documentation~\cite{tdxwhitepaper,tdxmodulespec,tdxarchspec,tdxghci,tdxloaderspec} and source code~\cite{tdxmodulesrc,tdxloadersrc,tdxkernelsrc}. 

The following is a roadmap of this paper. We begin by outlining the security principles (\S\ref{sec:principle}) and the threat model (\S\ref{sec:model}) of \ac{TDX}. Next, we provide a comprehensive comparison of existing confidential computing technologies on the market (\S\ref{sec:related}) and examine the existing Intel technologies that serve as the building blocks for \ac{TDX} (\S\ref{sec:background}). Once the background knowledge is established, we offer a high-level overview of \ac{TDX} (\S\ref{sec:overview}) and then delve into the technical details of the \ac{TDX} Module (\S\ref{sec:tdxmodule}), memory protection mechanisms (\S\ref{sec:memprotect}), and remote attestation (\S\ref{sec:remote_attestation}). Finally, we conclude with a summary (\S\ref{sec:conclusion}). To assist readers in navigating the numerous terms and abbreviations used in this paper, a list of acronyms is also provided (\S\ref{sec:acronym}).
\section{Security Principles} 
\label{sec:principle}
In cloud computing, multiple security domains, \eg a hypervisor managed by a cloud service provider and \acp{VM} owned by different tenants, coexist on a shared physical machine. While hardware-assisted virtualization can isolate tenants' workloads, the security model still relies on a privileged hypervisor to provide trustworthy \ac{VM} management. 
To address this issue, \ac{TDX} enforces cryptographic isolation among the security domains, thereby mitigating cross-domain attacks. This eliminates hierarchical dependencies on untrusted/privileged host software and excludes the hypervisor and cloud operators from the \ac{TCB}, allowing tenants to securely provision and run their computations with confidence.

\label{sec:principle:guarantees}
\ac{TDX} guarantees confidentiality and integrity of \ac{TD}'s memory and virtual CPU states,  ensuring that they cannot be accessed or tampered with by other security domains executing on the same machine. This is achieved through a combination of: (1) memory access control, (2) runtime memory encryption, and (3) an Intel-signed \ac{TDX} Module that handles security-sensitive \ac{TD} management operations. 

In addition, remote attestation provides tenants with proof of the authenticity of \acp{TD} executing on genuine \ac{TDX}-enabled Intel processors. These guarantees are based on a specific threat model and require certain trust assumptions, as described in \S\ref{sec:model}.

\nip{Memory Confidentiality.} \ac{TD}'s data residing inside the processor package are stored in clear text. However, when the data is offloaded from the processor to the main memory, the processor encrypts it using a TD-specific cryptographic key known only to the processor. The encryption is performed at the cache line granularity (as described in \S\ref{memoryprotection:mktme}), making it impossible for peripheral devices to read or tamper with the \ac{TD}'s private memory without detection. The processor is able to detect any tampering that may occur when loading data from the main memory.

\nip{CPU State Confidentiality.} \ac{TDX} protects against concurrently executing processes by managing the virtual CPU states of \acp{TD} during all context switches between security domains. The states are stored in the \ac{TD}'s metadata, which are protected while being in the main memory using the \ac{TD}'s key. During context switches, \ac{TDX} clears or isolates the \ac{TD}-specific states from internal processor registers and buffers, such as \ac{TLB} entries or branch prediction buffers, to maintain the protection of the \ac{TD}'s information.

\nip{Execution Integrity.} \ac{TDX} protects the integrity of \ac{TD}'s execution from host interference, ensuring that the \ac{TD} resumes its computation after an interrupt at the expected instruction within the expected states. It is capable of detecting malicious changes in the virtual CPU states, as well as injection, modification, or removal of instructions located in the private memory. However, \ac{TDX} does not provide additional guarantees for the control flow integrity. It is the responsibility of the \ac{TD} owner to use existing compilation-based or hardware-assisted control flow integrity enforcement techniques, such as \ac{CET}~\cite{cetwhitepaper}.

\nip{I/O Protection.} Peripheral devices or accelerators are outside the trust boundaries of \acp{TD} and should not be allowed to access \ac{TD}'s private memory. To support virtualized I/O, a \ac{TD} can choose to explicitly share memory for data transfer purposes. However, \ac{TDX} does not provide any confidentiality and integrity protection for the data located in shared memory regions. It is the responsibility of \ac{TD} owners to implement proper mechanisms, such as using secure communication channels like \ac{TLS}, to protect the data that leaves the \ac{TD}'s trust boundary. In the future, \ac{TDX} 2.0 is planned to include trusted I/O virtualization~\cite{deviceattestation,tdxio} to address these issues. 
\section{Threat Model}
\label{sec:model}
\ac{TDX} operates on the assumption that adversaries may have physical or remote access to a computer, and may be able to gain control over the boot firmware, \ac{SMM}, host operating system, hypervisor, and peripheral devices. The primary objective of these adversaries is to obtain confidential data or interfere with the execution of a \ac{TD}. It is important to note that \ac{TDX} cannot guarantee availability, as adversaries can control all the compute resources for \acp{TD} and launch \ac{DoS} attacks. It is crucial for the \ac{TDX} design to prevent adversaries from conducting actions that compromise the \ac{TDX} security guarantees outlined in \S\ref{sec:principle:guarantees}. Below, we summarize the capabilities of adversaries and identify potential attack vectors and scenarios.

Adversaries can interact with the \ac{TDX} Module through its host-side interface functions, which allow them to build, initialize, measure, and tear down \acp{TD}. These interface functions can be invoked in an arbitrary order with semantically and syntax valid/invalid inputs. 

Adversaries can control the compute resources assigned to \acp{TD}, including physical memory pages, processor time, and physical/virtual devices. They can interrupt \acp{TD} at any point, and try to read and write to arbitrary memory locations, as well as reconfigure the \ac{IOMMU}. 

Adversaries have the capability of manipulating the input data for \acp{TD}~\cite{tdxkernelhardening}, including \ac{ACPI} tables, \ac{PCI} config, \ac{MSR}, \ac{MMIO}, \ac{DMA}, emulated devices, hypercalls handled by the host, source of randomness, and time notion. 

Adversaries can conduct physical and hardware attacks, for instance, by probing buses or accessing main memory through malicious \ac{DMA}. There is no defense against physical attacks that roll back arbitrary memory regions. However, it should not be possible for adversaries to extract the secret key material baked into the processor chip's fuses. The scope of the threat model does not cover fault injections or side-channel attacks such as power glitches, time and power analysis.

Attacking \ac{TDX} attestation is within the scope as it undermines the trust model and may enable adversaries to forge a counterfeit \ac{TEE} for collecting confidential information from tenants. 

\nip{\acf{TCB}.}
The \ac{TCB} of \ac{TDX} consists of the TDX-enabled Intel processors and the built-in technologies, such as \ac{VT}, \ac{MKTME}, and \ac{SGX}. The \ac{TCB} also includes software modules signed by Intel, including the \ac{TDX} Module, NP/P-SEAM Loaders, and architectural \ac{SGX} enclaves for remote attestation. The software stacks running within \acp{TD} are owned by the tenant and are considered part of the \ac{TCB}. The cryptographic primitives used in \ac{TDX} are considered sound and its implementation secure, including the generation of random numbers and the absence of side-channel attacks like timing attack.

Tenants must trust the processor manufacturer, Intel, for developing, manufacturing, building, and signing of the hardware/software components used by \ac{TDX}. The source code packages for the \ac{TDX} Module, the NP/P-SEAM Loaders, and the \ac{DCAP} for attestation are publicly available for audit purposes, allowing tenants to assess their trustworthiness. However, tenants must also trust that the version signed by Intel is equivalent to the one they have reviewed, which involves placing trust in the compilation process to protect against supply chain attacks.

Moreover, tenants are required to trust Intel's \ac{PCS} for remote attestation. The \ac{PCS}, which originally supported \ac{SGX} attestation, has been expanded to include retrieval of \ac{PCK} certificates, revocation lists, and \ac{TCB} information for \ac{TDX}.
\section{Comparison of Confidential Computing Technologies} 
\label{sec:related}
Confidential computing technologies share a common objective of protecting outsourced sensitive data and computations from unauthorized access, tampering, and disclosure on untrusted third-party infrastructures. Major processor vendors are competing to incorporate confidential computing capabilities into their chips. Despite differences in implementation and terminology, these technologies share fundamental security principles with similar system designs, such as introducing new execution modes or privilege levels, migrating \ac{VM} management functions to attested firmware/software, ensuring secure or measured launch of trusted components, enforcing memory access control, and providing memory encryption protection.

In addition to Intel \ac{TDX}, we provide a brief overview of the confidential computing technologies from other vendors, including AMD \ac{SEV}, IBM Secure Execution and \ac{PEF}, Arm \ac{CCA}, and RISC-V \ac{AP-TEE}, for comparison purposes. We have summarized the distinct features of these technologies in Table~\ref{tab:summary-technologies-other}. Readers already familiar with these technologies can skip this section and proceed directly to \S\ref{sec:background}, where we explain the existing Intel technologies that support \ac{TDX}.

\begin{table}[t!]
\caption{Summary of Comparable Confidential Computing Technologies}
\label{tab:summary-technologies-other}
\small
\begin{tabular}{l|l}
\hline
\textbf{Technology} & \textbf{Summary} \\ \hline
AMD SEV~\cite{kaplan2016amd,kaplan2017protecting,sev2020strengthening} & \begin{tabular}[c]{@{}l@{}} - enforces cryptographic \ac{VM} isolation via AMD PSP\\ - supports memory encryption (\ac{SEV}), CPU state encryption (SEV-ES), integrity protection (SEV-SNP)\\
- provides hardware isolated layers within VMs through VMPL\end{tabular}\\ \hline
\begin{tabular}[c]{@{}l@{}}IBM Secure\\ Execution~\cite{ibmsecureexec}\end{tabular} & \begin{tabular}[c]{@{}l@{}}- protects SVMs on IBM Z and LinuxONE.\\ - leverages a trusted firmware, Ultravisor, to bootstrap and run SVMs \\ - provides end-to-end protection from the boot image to memory and throughout execution\end{tabular} \\ \hline
IBM PEF~\cite{hunt2021confidential} & \begin{tabular}[c]{@{}l@{}}- protects SVMs on Power ISA\\ - leverages the Protected Execution Ultravisor to manage SVM execution\\ - utilizes TPM, secure boot, and trusted boot for integrity check and bootstrap SVMs\end{tabular} \\ \hline
Arm CCA~\cite{li2022design} & \begin{tabular}[c]{@{}l@{}}- introduces Realm world for running confidential VMs\\ - introduces Root world to enforce address space isolation through GPT\\- support attestation for Realm environment\end{tabular} \\ \hline
RISC-V AP-TEE~\cite{riscvcc} & \begin{tabular}[c]{@{}l@{}} - introduces the TSM to manage TVM life-cycles \\ - uses MTT to track memory page assignment\\ - adopts a layered attestation architecture \end{tabular} \\\hline
\end{tabular}
\end{table}

\subsection{AMD SEV} 
\acf{SEV}~\cite{kaplan2016amd} is a confidential computing feature in AMD EPYC processors. It protects sensitive data stored within \acp{VM} from privileged software or administrators in a multi-tenant cloud environment.  \ac{SEV} relies on AMD \ac{SME} and AMD Virtualization (AMD-V) to enforce cryptographic isolation between \acp{VM} and the hypervisor. Each \ac{VM} is assigned a unique ephemeral \ac{AES} key, which is used for runtime memory encryption. The \ac{AES} engine in the on-die memory controller encrypts or decrypts data written to or read from the main memory. The per-\ac{VM} keys are managed by the AMD \ac{PSP}, which is a 32-bit Arm Cortex-A5 micro-controller integrated within the AMD \ac{SoC}. The C-bit (bit 47) in physical addresses determines memory page encryption. \ac{SEV} also provides a remote attestation mechanism that allows the \ac{VM} owners to verify the trustworthiness of \acp{VM}' launch measurements and the \ac{SEV} platforms. The \ac{PSP} generates the attestation report signed by an AMD certified attestation key. The \ac{VM} owners can verify the authenticity of the attestation report and the embedded platform/guest measurements.

AMD has released three generations of \ac{SEV}. The first generation \ac{SEV}\cite{kaplan2016amd} only protects the confidentiality of a \ac{VM}'s memory. The second generation SEV-ES (Encrypted State)\cite{kaplan2017protecting} adds protection for CPU register state during hypervisor transition, and the third generation SEV-SNP (Secure Nested Paging)\cite{sev2020strengthening} adds integrity protection to prevent memory corrupting, replaying, and remapping attacks. Particularly, SEV-SNP provides memory integrity protection using \acf{RMP}. \ac{RMP} tracks each page's ownership and permissions to prevent unauthorized access. SEV-SNP also introduces the \acp{VMPL} feature by dividing the guest address space into four levels and providing additional security isolation within a \ac{VM}. The privilege levels range from zero to three, where \ac{VMPL}0 is the highest level of privilege and \ac{VMPL}3 is the lowest. For instance, the Linux Secure \ac{VM} Service Module (SVSM)~\cite{linuxsvsm} makes extensive use of the \ac{RMP} and \ac{VMPL} features to perform sensitive services, \eg live migration and vTPM, in a secure manner. 

\subsection{IBM Confidential Computing}
IBM's early exploration of confidential computing can be traced back to the research on SecureBlue++~\cite{williams2011cpu, boivie2012secureblue++}, which included running on an emulated POWER processor on the Mambo CPU simulator~\cite{bohrer2004mambo}. Today, IBM Systems support two architectures for confidential computing: Secure Execution~\cite{ibmsecureexec}, offered on IBM Z and LinuxONE, and \acf{PEF}~\cite{hunt2021confidential,pef}, released as an open source project on OpenPOWER systems.

\nip{IBM Secure Execution.}
IBM Secure Execution provides support for \acp{SVM} that run inside isolated \acp{TEE} since IBM Z15 and LinuxONE III. Secure Execution protects the confidentiality, integrity and authenticity of code and data in an \ac{SVM} from any unauthorized access and snooping or tampering. Secure Execution leverages trusted firmware, called the \spec{Ultravisor}, to perform security-sensitive tasks to bootstrap and run \acp{SVM}. The Ultravisor shields the \ac{SVM}'s memory and its state during context switches and protects the \ac{SVM} from a potentially compromised or malicious hypervisor. Tenants using Secure Execution can embed their encrypted sensitive data in the \ac{VM} images and rely on the Ultravisor to decrypt and expose them to the \acp{SVM} executing inside the \acp{TEE}. Specifically, tenants can encrypt their confidential data with a symmetric \spec{data key}, which they embed in the \spec{IBM Secure Execution header}. They further encrypt this header with the key obtained from the verified \spec{host key document} and embed the header in their \ac{VM} image. The header can contain multiple key slots that allow an image to run on multiple target hosts. The host key document, signed by the hardware manufacturer, contains the public key linked with the private key embedded in the hardware of IBM Z or LinuxONE. Ultravisor, the only component having access to the hardware private key and the data key, enforces that only the expected tenant's \ac{SVM} executing inside the \ac{TEE} has access to the unencrypted data. In addition to embedding built-in secrets within the \ac{VM} image, Secure Execution also supports remote attestation starting from IBM Z16 and LinuxONE Emperor 4. This allows tenants to verify the \ac{SVM}'s measurements before releasing their secrets.

\nip{IBM \ac{PEF}.} 
\ac{PEF} provides a \ac{VM}-based \ac{TEE} using extensions to the IBM Power \ac{ISA} that are supported in most POWER9 and POWER10 processors. \ac{PEF} firmware, tooling to prepare \acp{SVM}, and OS extensions, were released as open source software~\cite{ultravisorsrc}. To protect sensitive data and code, \ac{PEF} introduces a trusted firmware called \spec{Protected Execution Ultravisor} (Ultravisor) that shields the \ac{SVM} execution and enforces the security guarantees with the help of the CPU architectural changes. 
The \ac{PEF} relies on secure and trusted boot of the system and the Ultravisor executing in a new, highest privileged CPU state called \emph{secure state}. The hypervisor starts the \ac{VM}, which invokes the Ultravisor to transition to an \ac{SVM} using the \ac{ESM} call. The Ultravisor converts the \ac{VM} into an \ac{SVM} by moving it to the secure memory that is  inaccessible to untrusted code. Before executing the \ac{SVM}, the Ultravisor performs integrity checking. It decrypts the payload attached to the \ac{SVM} image to decode the integrity information and a passphrase for the encrypted file system.  After ensuring the integrity of the \ac{SVM}, the Ultravisor exposes the passphrase to the \ac{SVM} booting system that decrypts the tenant's file system. The Ultravisor uses the \ac{TPM} to get access to the symmetric seed required to check integrity and decrypt the payload. The symmetric seed is guarded using the \ac{PCR} sealing mechanism and accessed by establishing a secure channel to the \ac{TPM}. The \ac{TPM} only grants access for an Ultravisor on a correctly booted system. If the Ultravisor gets access to the symmetric seed, it generates the HMAC key and symmetric key that are used to verify integrity and decrypt the passphrase.

\subsection{Arm CCA} 
\acf{CCA}~\cite{li2022design} was introduced in the Armv9 architecture. Traditionally, Arm TrustZone allows secure execution by having two separated worlds, the \spec{Normal World} and the \spec{Secure World}. TrustZone prevents software in \spec{Normal World} from accessing data in the \spec{Secure World}. \ac{CCA} introduces the \ac{RME} with two additional worlds, the \spec{Realm World} and the \spec{Root World}. The \spec{Realm World} provides mutually distrusting execution environments for confidential \acp{VM}, isolating workloads from any other security domains, including host operating systems, hypervisors, other Realms and the TrustZone. 
To enforce isolation of address spaces, \ac{CCA} uses a \ac{GPT}, which is an extension to the page table that tracks the ownership of each page with different worlds. The \spec{Monitor} in the \spec{Root World} handles the creation and management of the \ac{GPT}, preventing hypervisors or operating systems from directly changing it. The \spec{Monitor} can dynamically move physical memory between different worlds by updating the \ac{GPT}. \ac{CCA} also supports attestation to measure and verify the \ac{CCA} platform and the initial state of the Realms.

\subsection{RISC-V AP-TEE}
\acf{AP-TEE}~\cite{riscvcc} is a reference confidential computing architecture for RISC-V. Its protected instance is called a \ac{TVM}. The architecture introduces the \ac{TSM} driver, which is a M-mode (highest privilege level in RISC-V) firmware component for switching between confidential and non-confidential environments. The \ac{TSM} driver tracks the assignment of memory pages to \acp{TVM} through the \ac{MTT}. The \ac{TSM} driver measures and loads the \ac{TSM}, which is a trusted intermediary between the hypervisor and the \acp{TVM}. \ac{AP-TEE} defines the \ac{ABI} for the hypervisor to request virtual machine management services from the \ac{TSM}. \ac{AP-TEE} adopts a layered attestation architecture, which begins with the hardware and progresses through the \ac{TSM} driver, \ac{TSM}, and \ac{TVM}. Each layer is loaded, measured, and certified by the previous layer. This approach provides a secure chain of trust that can be used to verify the integrity of the system. The \ac{TVM} can obtain a certificate from the \ac{TSM} that contains attestation evidence rooted back to the hardware. This certificate provides a mechanism for verifying the authenticity of the \ac{TVM} and the software it runs.
\section{Building Blocks for \acs{TDX}}
\label{sec:background}
\ac{TDX} relies on a combination of existing Intel technologies, including \acf{VT}, \acf{TME}/\acf{MKTME}, and \acf{SGX}. In this section, we provide an overview of these underpinning technologies and explain how they are used in \ac{TDX}. A summary of these technologies can be found in Table~\ref{tab:summary-technologies-intel}.

\begin{table}[t!]
\caption{Summary of Existing Building Blocks for TDX}
\label{tab:summary-technologies-intel}
\small
\begin{tabular}{l|l}
\hline
\textbf{Technology} & \textbf{Summary} \\ \hline
Intel VT~\cite{vtx} & \begin{tabular}[c]{@{}l@{}}- provides hardware-assisted virtualization for CPU, memory, and I/O\\
- enforces isolation among \acp{VM} via a trusted hypervisor
\end{tabular}\\ \hline
Intel TME & \begin{tabular}[c]{@{}l@{}}- encrypts entire main memory\\ - uses a single and boot-time generated transient key\\ - uses the AES-XTS algorithm with 128-bit keys or 256-bit keys\end{tabular} \\ \hline
Intel MKTME~\cite{mktmewhitepaper} & \begin{tabular}[c]{@{}l@{}}- supports multiple keys for memory encryption\\ - enables memory encryption at the page granularity\end{tabular} \\ \hline
Intel SGX~\cite{mckeen2013innovative} & \begin{tabular}[c]{@{}l@{}}- encloses sensitive code and data of an application within an enclave\\ - protects against memory bus snooping and cold boot attacks with memory encryption\\ - supports local and remote attestation\end{tabular} \\ \hline
\end{tabular}
\end{table}

\subsection{Intel VT}
Intel \ac{VT}~\cite{vtx} is a set of hardware-assisted virtualization features in Intel processors. Using \ac{VT}, \acp{VMM} or hypervisors can achieve better performance, isolation, and security compared to the software-based virtualization. Intel's \ac{VT} portfolio includes, among others, virtualization of CPU, memory, and I/O. 

Processors with \ac{VT}-x technology have a special instruction set, called \ac{VMX}, which enables control of virtualization. Processors with VT-x technology can operate in two modes: \spec{\ac{VMX} root mode} and \spec{\ac{VMX} non-root mode}. The hypervisor runs in \spec{\ac{VMX} root mode} while the guest \acp{VM} run in the \spec{\ac{VMX} non-root mode}. \ac{VT}-x defines two new transitions, \ac{VM} entry and \ac{VM} exit, to switch between the guest and the hypervisor. The \ac{VMCS} is a data structure that stores \ac{VM} and host state information for mode transitions. It also controls which guest operations can cause \ac{VM} exits.

Intel \ac{VT}-x utilizes \ac{EPT} for implementing \ac{SLAT}. Each guest kernel maintains its own page table to translate \ac{GVA} to \ac{GPA}. The hypervisor manages \ac{EPT} to map \ac{GPA} to \ac{HPA}. 

\acp{VM} can use different I/O models, including software-based and hardware-based models, to access I/O devices. Software-based I/O models involve emulated devices or para-virtualized devices, while hardware-based I/O models include direct device assignment, \ac{SR-IOV} devices, and \ac{S-IOV} devices.

Intel \ac{VT} for Directed I/O (\ac{VT}-d) enables the isolation and restriction of device accesses to entities managing the device. It includes I/O device assignment, \ac{DMA} remapping, interrupt remapping, and interrupt posting. With the support of \ac{VT}-d, \acp{VM} can directly access physical I/O memory through virtual-to-physical address translation with the help of the \ac{IOMMU}. \ac{VT}-d also provides flexibility in I/O device assignments to \acp{VM} and eliminates the need for the hypervisor to handle interrupts and \ac{DMA} transfers. Overall, \ac{VT}-d enhances the performance and security of virtualized environments that require direct access to I/O devices.

\nip{\ac{VT} $\Rightarrow$ \ac{TDX}.} \ac{TDX} is a \ac{VM}-based \ac{TEE}. It relies on the \ac{VT} to provide isolation among \acp{TD}. As the hypervisor is no longer trusted in the new threat model, the functionalities of managing \acp{TD} have been enclosed within the \ac{TDX} Module. The \ac{TDX} Module and \acp{TD}  run in the new \spec{\ac{SEAM} \ac{VMX} root/non-root mode} with additional protection. \ac{TDX} still leverages \ac{EPT} to manage \ac{GPA}-to-\ac{HPA} translation. But currently it maintains two \acp{EPT} for each \ac{TD}, a protected one for private (encrypted) memory and another one for shared (unencrypted) memory. We explain in more details the \ac{TDX}'s architecture and the \ac{TDX} Module in \S\ref{sec:overview:architecture} and \S\ref{sec:tdxmodule}. 

It is worth noting that currently nested virtualization is not supported in \ac{TDX} 1.0, which means that running \acp{VM} within a \ac{TD} is not allowed. Attempting to use \ac{VMX} instructions within a \ac{TD} can result in \ac{UD} exceptions. But the TD partitioning architecture specification draft~\cite{tdpartitioning} indicates that nested virtualization will be supported in \ac{TDX} 1.5 in the future.   

\subsection{Intel TME/MKTME}
\label{sec:background:mktme}
\ac{TME} was first introduced with the Intel 11th Generation Core vPro mobile processor. This feature is designed to protect against attackers who have physical access to a computer's memory and attempt to steal data. \ac{TME} encrypts the entire computer's memory using a single transient key. The key is generated at boot-time through a combination of hardware-based random number generators and security measures integrated into the system's chipset. Memory encryption is performed by encryption engines on each memory controller. The encryption process uses the NIST standard AES-XTS algorithm with 128-bit or 256-bit keys.

\ac{MKTME}~\cite{mktmewhitepaper} extends \ac{TME} to support multiple keys and memory encryption at page granularity. For each memory transaction, \ac{MKTME} selects an encryption key to encrypt memory based on the \ac{HKID}. \ac{HKID} occupies a configurable number of bits from the top of the physical address. The \acp{HKID} range is set by the BIOS during system boot. \ac{MKTME} allows for software-provided keys and introduces a new instruction, \code{PCONFIG}, for programming the key and encryption mode associated with a particular \ac{HKID}. These \code{$\langle\ac{HKID}, key\rangle$} tuples are stored in the \ac{KET}, which is held by each \ac{MKTME} encryption engine. The keys in the \ac{KET} never leave the processor and are never exposed to software. \ac{MKTME} can be used in both native and virtualized environments. In the virtualized environments, hypervisors control the memory encryption for different \acp{VM} by attaching \acp{HKID} to \ac{VM}'s physical addresses in \ac{EPT}.  

\nip{\ac{MKTME} $\Rightarrow$ \ac{TDX}.} To use \ac{MKTME} in the virtualized environments, the hypervisor must be trusted to control the memory encryption, which violates the new threat model for confidential computing. Therefore, in \ac{TDX}, the \ac{TDX} Module is responsible for controlling memory encryption for \acp{TD}. The \ac{HKID} space has been partitioned to private \acp{HKID} and shared \acp{HKID}. The private \acp{HKID} can only be used for encrypting private memory of \acp{TD}. The \ac{TDX} Module still leverages \ac{MKTME} to protect \ac{TD}'s memory. More information about how \ac{TDX} uses \ac{MKTME} can be found in \S\ref{sec:overview:mem_protect} and \S\ref{sec:memprotect}.

\subsection{Intel SGX}
Intel introduced \ac{SGX}~\cite{mckeen2013innovative} in 2015 with the 6th Generation Core processors to protect against memory bus snooping and cold boot attacks. It enables developers to partition their applications and protect selected code and data within enclaves. The memory of an enclave can only be accessed by authorized code. \ac{SGX} uses hardware-based memory encryption to protect the enclave's contents. Any unauthorized attempts to access or tamper with the enclave's memory can trigger exceptions.
\ac{SGX} adds 18 new instructions into Intel's \ac{ISA} and enables secure offloading of computations to environments where the underlying host components (such as hosting application, host kernel, \ac{SMM}, and peripheral devices) are untrustworthy. \ac{SGX}'s security ultimately depends on the security of the firmware and microcode that implement its features.

The \ac{EPC} is a special memory region that contains the enclave's code and data, where each page is encrypted using the \ac{MEE}. The \ac{EPCM} stores the page metadata, such as configuration, permissions, and type of each page. 
At boot time, keys are generated and used for decrypting the contents of encrypted pages inside the CPU. The keys are controlled by the \ac{MEE} and never exposed to the outside. Thus, only this particular CPU can decrypt the memory.
The CPU stores these keys internally and prevents access to them by any software. Additionally, privileged software out of enclaves is not allowed to read or write the \ac{EPC} or \ac{EPCM} pages.

\ac{SGX} offers both local and remote attestation to verify the integrity and authenticity of enclaves. Local attestation is used to establish trust between two enclaves within the same platform, while remote attestation verifies the trustworthiness of an enclave to a third-party entity off the platform.
In local attestation, an enclave can verify another enclave's integrity and the genuineness of the underlying hardware platform. 
To do so, the first enclave generates a \spec{report} and uses the identity information of the second enclave to sign it. The second enclave retrieves its Report Key and verifies the report using this Report Key. 
A third party may want to establish trust with a remotely executed enclave before provisioning it with secrets.
In this scenario, remote attestation is necessary. To perform remote attestation, \ac{SGX} utilizes a special architectural enclave known as the \ac{QE}. The \ac{QE} is developed and signed by Intel. 
The \ac{QE} receives a \spec{report} from another enclave, locally verifies it, and transforms it into a remotely verifiable \spec{quote} by signing it with the Attestation Key. The relying party can send this \spec{quote} to the \ac{IAS}, which verifies the \spec{quote} to identify and assess the trustworthiness of the \ac{SGX} enclave. The \ac{QE}'s role is to provide a secure and trustworthy environment for the transformation of a \spec{report} into a \spec{quote}, and to ensure the \spec{quote} cannot be modified or falsified. Intel also provides \acf{DCAP}~\cite{dcapwhitepaper}, which is a composition of software packages, for datacenters to deploy their own ECDSA attestation infrastructures for \ac{SGX} enclave attestation.

Researchers have used \ac{SGX} to provide secure containers (\eg SCONE~\cite{arnautov2016scone}) and shielded execution for unmodified applications (\eg Haven~\cite{baumann2015shielding}). Graphene~\cite{tsai2017graphene}, an \ac{SGX}-based framework, provides techniques for running unmodified applications as well as dynamic libraries inside \ac{SGX} enclaves. Besides, \ac{SGX} has a wide spectrum of applications ranging from the function encryption system (\eg Iron~\cite{fisch2017iron}), source code partitioning to protect security-sensitive data and functions (\eg Glamdring~\cite{lind2017glamdring}), machine learning~\cite{ohrimenko2016oblivious,gu2018confidential,tramer2018slalom,gu2019reaching}, network security~\cite{beekman2016attestation}, fault-tolerant~\cite{behl2017hybrids}, encrypted data search (\eg HardIDX~\cite{fuhry2017hardidx}), secure databases (\eg EnclaveDB~\cite{priebe2018enclavedb}), secure coordination for distributed system (\eg SecureKeeper~\cite{brenner2016securekeeper}), and secure distributed computations (\eg VC3~\cite{schuster2015vc3}). 
Identifying vulnerabilities of \ac{SGX} is another important line of research. Researchers also have identified a wide range of attack vectors targeting \ac{SGX}, such as controlled-channel attacks~\cite{xu2015controlled, van2017telling, van2017sgx, wang2017leaky}, cache attacks~\cite{moghimi2017cachezoom, gotzfried2017cache, schwarz2017malware, brasser2017software}, branch prediction attacks~\cite{lee2017inferring, evtyushkin2018branchscope}, and speculative execution attacks~\cite{chen2019sgxpectre, koruyeh2018spectre}. 

\nip{\ac{SGX} $\Rightarrow$ \ac{TDX}.} \ac{SGX} and \ac{TDX} protect memory at different granularities. But on the same platform, \ac{TDX} and \ac{SGX} are within the same \ac{TCB}. Thus, they can locally attest each other. \ac{TDX} leverages the remote attestation mechanism provided by \ac{SGX}. The attestation report of a \ac{TDX} platform can be verified and signed within a \acl{QE}. More details about \ac{TDX}'s remote attestation can be found in \S\ref{sec:overview:attestation} and \S\ref{sec:remote_attestation}. 

It is worth noting that at the moment, running an \ac{SGX} enclave within a \ac{TD} is not allowed, as invoking \code{ENCLS}/\code{ENCLV} instructions within a TD can lead to \ac{UD} exceptions.
\section{Overview of TDX}
\label{sec:overview}
In this section, we give an overview of \ac{TDX}, discussing its system architecture, memory protection mechanisms, I/O model, attestation, and features that have been planned for the future. Each topic also includes pointers to subsequent sections that provide more technical details.  

\begin{figure}[t!]
\centerline{\includegraphics[width=0.8\textwidth]{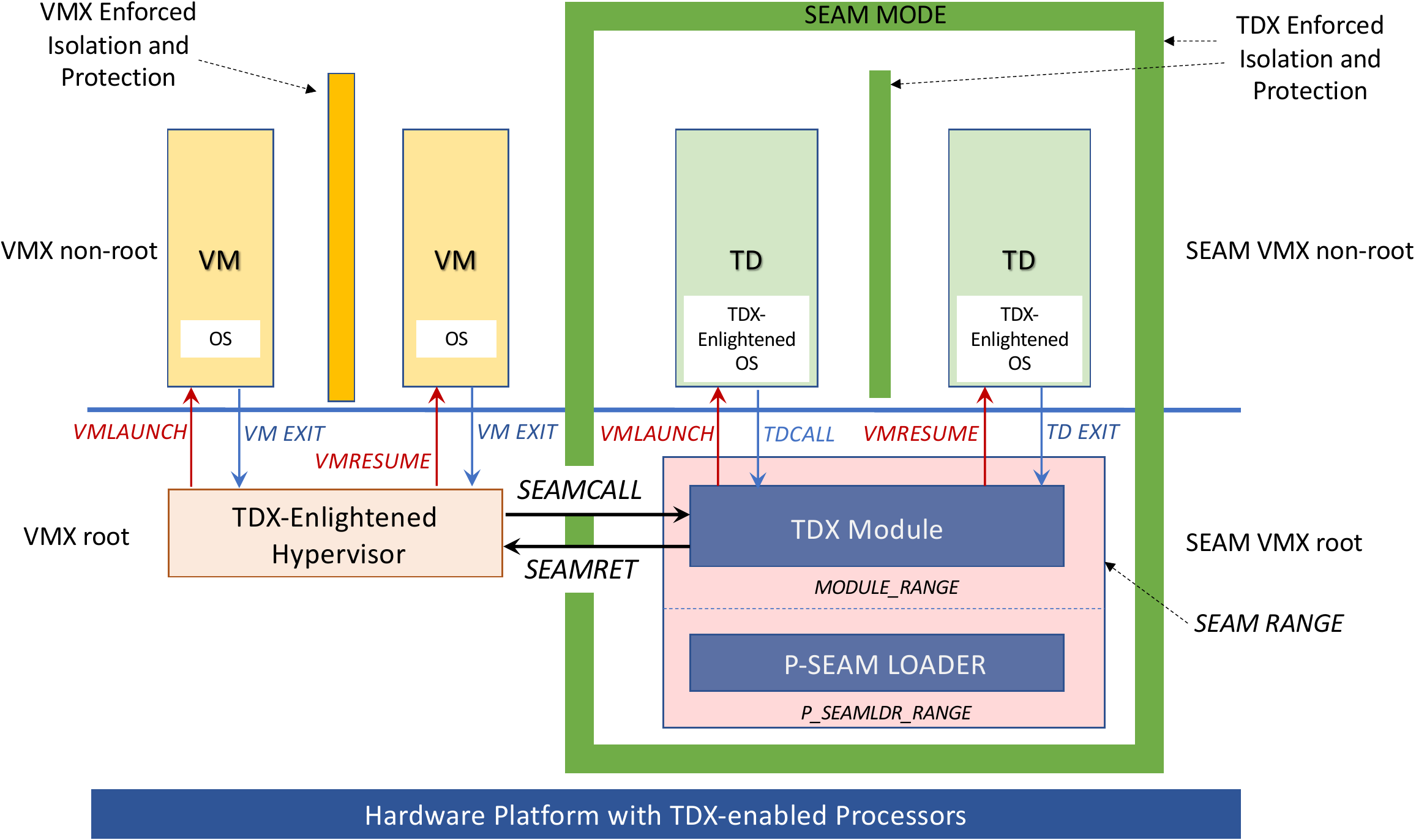}}
\caption{TDX System Architecture}
\label{fig:tdx_arch_overview}
\end{figure}
\subsection{TDX System Architecture}
\label{sec:overview:architecture}
\autoref{fig:tdx_arch_overview} illustrates the runtime architecture of \ac{TDX}. It is composed of two key components: (1) TDX-enabled processors, which offer architectural functionalities like hardware-assisted virtualization, memory encryption/integrity protection, and the ability to certify \ac{TEE} platforms, (2) \ac{TDX} Module, an Intel-signed and CPU-attested software module that leverages the features of \ac{TDX}-enabled processors to facilitate the construction, execution, and termination of \acp{TD} while enforcing the security guarantees. The \ac{TDX} Module provides two sets of interface functions, host-side interface functions for a \ac{TDX}-enlightened hypervisor and guest-side interface functions for \acp{TD}. It is loaded and executed in the \code{\ac{SEAM} RANGE}, which is a portion of system memory reserved via UEFI/BIOS. The P-SEAM Loader, which also resides in the \code{\ac{SEAM} RANGE}, can install and update the \ac{TDX} Module. More information on the loading process of the \ac{TDX} Module can be found in \S\ref{sec:tdxmodule:load}.

\acf{SEAM} is an extension to the \ac{VMX} architecture and provides two new execution modes: \spec{\ac{SEAM} \ac{VMX} root mode} and \spec{\ac{SEAM} \ac{VMX} non-root mode}. A \ac{TDX}-enlightened hypervisor operates in the traditional \spec{\ac{VMX} root mode} and utilizes the \code{SEAMCALL} instruction to call host-side interface functions (function names start with \code{TDH}) of the \ac{TDX} Module. Upon execution of the \code{SEAMCALL} instruction, the logical processor transitions from the \spec{\ac{VMX} root mode} into \spec{\ac{SEAM} \ac{VMX} root mode} and starts executing code within the \ac{TDX} Module. Once the \ac{TDX} Module has completed its task, it returns to the hypervisor in \spec{\ac{VMX} root mode} by executing the \code{SEAMRET} instruction. On the other hand, \acp{TD} run in the \spec{\ac{SEAM} \ac{VMX} non-root mode}. \acp{TD} can trap into the \ac{TDX} Module either through a \ac{TD} exit or by invoking the \code{TDCALL} instruction. In both cases, the logical processor transitions from the \spec{\ac{SEAM} \ac{VMX} non-root mode} into the \spec{\ac{SEAM} \ac{VMX} root mode} and starts executing in the context of the \ac{TDX} Module. The names of guest-side interface functions handling \code{TDCALLs} start with \code{TDG}. Details about the \ac{TDX} context switches can be found in \S\ref{sec:tdxmodule:context}.

\subsection{TDX Memory Protection}
\label{sec:overview:mem_protect}
\ac{TDX} leverages \ac{VMX} to enforce the memory isolation for \acp{TD}. Similar to legacy \acp{VM}, \acp{TD} are unable to access the memory of other security domains, such as \ac{SMM}, hypervisors, the \ac{TDX} Module, and other \acp{VM}/\acp{TD}. With \ac{VMX}, hypervisors maintain \acp{EPT} to enforce memory isolation. However, since hypervisors are no longer trusted, \ac{TDX} has moved the task of memory management to the \ac{TDX} Module, which controls the address translation of \ac{TD}'s private memory. 

A more intriguing aspect of \ac{TDX}'s security model is its protection of \ac{TD}'s memory from privileged software, corrupted devices, and unprincipled administrators on the host. \ac{TDX} achieves this by implementing \spec{access control} and \spec{cryptographic isolation}. Access control prevents other security domains on the same computer from accessing a \ac{TD}'s data. Cryptographic isolation is utilized to prevent malicious \ac{DMA} devices or adversaries with physical access to the main memory from directly reading or corrupting \ac{TD}'s memory.

\nip{Memory Partitioning.}
With \ac{TDX} enabled, the entire physical memory space is partitioned into two parts: \spec{normal memory} and \spec{secure memory}. The sensitive data of \acp{TD}, including the private memory, virtual CPU state, and its associated metadata, should be stored in secure memory. \acp{TD} can also specify memory regions as shared memory for I/O, which is not protected through \ac{TDX}, and thus belong to normal memory. All other software, which is \emph{not} executing in the SEAM mode, belongs to normal memory and is not allowed to access secure memory, regardless of its privilege level. The memory controller, an architectural component inside the processor, enforces memory access checks.

To make a physical page part of the secure memory, the \spec{\ac{TD} Owner bit} is enabled (\S\ref{sec:memprotect:TD_memory_integrity}). Each \spec{\ac{TD} Owner bit} is associated with a memory segment corresponding to a cache line\footnote{At the time of this writing, the size of the processor's cache line is 64-byte; thus the address of such a memory segment is 64-byte aligned.}. The \spec{\ac{TD} Owner bits} are stored in the \ac{ECC} memory associated with these segments. The \ac{TDX} Module controls the conversion of physical memory pages to secure memory by attaching private \acp{HKID} to their physical addresses. The \ac{HKID} is encoded in the upper bits of the physical address. The set of private \acp{HKID} are controlled by \ac{TDX} and can only be used for \acp{TD} and the \ac{TDX} Module. When the memory controller writes to a physical address with a private \ac{HKID}, it sets the \spec{\ac{TD} Owner bit} to $1$. When it writes to an address that does not have a private \ac{HKID}, it clears the \spec{\ac{TD} Owner bit}. Access control is enforced on each cache line read. The read request passes through the memory controller, which permits only processes executing in \ac{SEAM} mode to read a cache line with a \spec{\ac{TD} Owner bit} set to $1$. Any read request not in the \ac{SEAM} mode receives all zeros when trying to read such a cache line.

When building a \ac{TD}, the (untrusted) hypervisor selects the memory pages from the normal memory to become part of the secure memory. The \ac{TDX} Module gradually moves these pages to the secure memory. It uses them for the metadata (\S\ref{sec:tdxmodule:td_metadata}) and the main memory of each individual \ac{TD}. A TD must explicitly accept these pages before they can be used for its main memory. The \ac{TDX} Module performs sanity checks of the secure memory setup by maintaining a \ac{PAMT}, which is described in more detail in \S\ref{sec:tdxmodule:mem_management}. 

\nip{Memory Confidentiality.}
\label{memoryprotection:mktme}
\ac{TDX} leverages \ac{MKTME} (\S\ref{sec:background:mktme}) for encrypting \ac{TD}'s private memory and its metadata. \ac{MKTME} is responsible for transparent memory encryption and decryption of data passing through the memory controller. The \ac{TDX} Module programs the keys used by the \ac{MKTME} to encrypt specific cache lines when they are written to memory. The keys are associated with the \acp{HKID} embedded in the physical addresses. \ac{MKTME} decodes \acp{HKID} and uses the referenced cryptographic keys to perform the cryptographic operations. 

\ac{MKTME} stores cryptographic keys in its internal memory, never exposing them to the outside. The cryptographic keys can only be referenced by their \acp{HKID}. When building a new TD, the hypervisor selects an unused private \ac{HKID}, and the \ac{TDX} Module requests the processor to generate a new cryptographic key related to this \ac{HKID}. The \ac{TDX} Module binds this key to the \ac{TD}. It guarantees that the memory of each TD is encrypted with a different cryptographic key. 

\ac{MKTME} encrypts memory at the cache line granularity using AES-128 XTS cryptography. Each cache line in the main memory is associated with metadata that encodes its association with secure memory (via a \spec{\ac{TD} Owner bit}) and, optionally, a \ac{MAC} for integrity and authentication checking. The encryption can prevent some physical attacks, like the cold boot attack. Please see \S\ref{sec:memprotect:MKTME} for more details on \ac{MKTME} and \acp{HKID}.

\nip{Memory Integrity.}
\label{sec:overview:mem_integrity}
\ac{TDX} provides two distinct mechanisms for ensuring memory integrity: \spec{\ac{Li}} and \spec{\ac{Ci}}. \ac{Li} protects the integrity against software-level attacks by preventing unauthorized encryption and reads of secure memory. It restricts the use of private \acp{HKID} and reads of secure memory. It checks that only processes running in the \ac{SEAM} mode can encrypt with the private \ac{HKID} and read a cache line whose \spec{\ac{TD} Owner bit} is set to $1$. This functionality does not prevent adversaries with direct access to the main memory from setting arbitrary content of the cache line and \spec{\ac{TD} Owner bit}. This opens up a possibility for rollbacks of individual cache lines. 
\ac{Ci} is an advanced mechanism that addresses the limitations of \ac{Li}, which does not detect malicious direct memory writes or bit flips (\eg via a Rowhammer attack~\cite{rowhammer-ieee-2014}). It prevents an adversary with physical access to the main memory from tampering with the memory content. We provide a more detailed technical discussion of the memory integrity protection in \S\ref{sec:memprotect:TD_memory_integrity}.

\subsection{TDX I/O Model}
According to the \ac{TDX} threat model, hypervisors and peripheral devices are considered untrusted and are prohibited from directly accessing the private memory of \acp{TD}. It is the responsibility of \acp{TD} and their owners to secure I/O data before it leaves the trust boundary. This requires sealing the I/O data buffers and placing them in shared memory, which is identified by the \spec{shared} bit in the \acl{GPA}. Hypervisors or peripheral devices can then move the data in and out of the shared memory. This necessitates modifications to the guest kernel to support this I/O model. Furthermore, all I/O data that is transferred into the \acp{TD} from hypervisors or peripheral devices must be thoroughly examined and validated, as it is no longer considered trustworthy.

In the Linux guest support for \ac{TDX}, all \ac{MMIO} regions and \ac{DMA} buffers have been mapped as shared memory within the \acp{TD}. The Linux guest is enforced to use \spec{SWIOTLB} to allocate and convert \ac{DMA} buffers in unified locations. To protect against malicious inputs from I/O, only a limited number of hardened drivers~\cite{tdxkernelhardening} are allowed within \acp{TD}.

\subsection{TDX Attestation}
\label{sec:overview:attestation}
Remote attestation is a method for verifying the identity and trustworthiness of a \ac{TEE}. The attester can provide proof to a challenger to show that  computations are being executed within protected domains. The challenger validates the evidence by checking the digital signatures and comparing the measurements to reference values.  

On a \ac{TDX}-enabled machine, the attester operates within a \ac{TD} and is responsible for handling remote attestation requests. When a request is received from a challenger, such as a tenant, the attester provides evidence of proper instantiation of the TD through the generation of a \ac{TD} \spec{quote}. This \spec{quote}, which serves as the evidence, is produced by the \ac{TDX} module and signed by the \acl{QE}. It contains measurements of the \ac{TDX}'s \ac{TCB} and the software components loaded in the \ac{TD}. The \spec{quote} also includes a certificate chain anchored by a certificate issued by Intel. Upon receipt of the \spec{quote}, the challenger verifies its authenticity by checking the \spec{quote} and determining if the attester is running on a genuine \ac{TDX}-enabled platform and if the \ac{TD} has the expected software measurements. If the \spec{quote} is successfully validated, the challenger can proceed to establish a secure channel with the attester or release secrets to the attester. We provide a more detailed technical discussion of remote attestation in \S\ref{sec:remote_attestation}.

\subsection{Future Features}
Live migration and \ac{TEE} I/O are crucial features for confidential \acp{VM} but are currently not supported in \ac{TDX} 1.0. However, according to documents~\cite{tdx-migration,tdxio,deviceattestation}, Intel is planning to include the support for live migration in \ac{TDX} 1.5 and TEE I/O in \ac{TDX} 2.0. These plans are still in progress and may subject to change in the future. Here we provide a brief overview of these two features and explain their design. 

\nip{Live Migration.}
Live migration is an essential feature for cloud service providers as it enables them to transfer running \acp{VM} from one physical host to another without any service interruptions. This functionality is important for maintenance tasks such as hardware upgrades, software patches, and load balancing. However, migrating a \ac{TD} is more complex than migrating a traditional \ac{VM} due to the security concerns of confidential computing. Since the hypervisor is considered untrusted, it is not allowed to directly access and transfer the CPU state and private memory of the \ac{TD} from the source to the destination platform. Furthermore, tenants should have the ability to define and enforce migration policies. For instance, if the destination platform does not meet the \ac{TCB} requirements specified in the policy, the migration should be cancelled. 

Intel introduces Service \acp{TD} to expand the trust boundary of the \ac{TDX} Module. Rather than making the \ac{TDX} Module overly complex and bloated, it is more convenient and flexible to add customized and specialized functionalities into a Service \ac{TD}. A Service TD can be bound to regular \acp{TD} via the \ac{TDX} Module with the access privileges to their assets.

\ac{MigTD} is a Service \ac{TD} that is specifically designed for live migration. The entire live migration session is under the control of the \ac{TDX} Module and the \acp{MigTD}. The untrusted hypervisor, which is controlled by the cloud service provider, is only responsible for transferring the encrypted \ac{TD}'s assets over networks. These assets include the \ac{TD}'s metadata, CPU state, and private memory, and are protected by a \ac{MSK} that is only accessible by the \acp{MigTD} and \ac{TDX} Module.

Both the source and destination platforms have a running \ac{MigTD}. \acp{MigTD} are bound to the source \ac{TD} (to be migrated) and the destination \ac{TD} (initially as a \ac{TD} template waiting for migration), respectively. The \acp{MigTD} are responsible for remote attestation between source and destination platforms and evaluate their \ac{TCB} levels based on security policies. Once the platforms are deemed acceptable for migration, a secure channel is established between the two \acp{MigTD}. The source \ac{MigTD} generates a \ac{MSK}, which is shared to the destination \ac{MigTD} through this secure channel. Both \acp{MigTD} program the \ac{MSK} into the corresponding \ac{TDX} Modules. The source \ac{TDX} Module exports and encrypts the \ac{TD}'s assets with the \ac{MSK}, while the destination \ac{TDX} Module decrypts the assets with the same key and imports them into the destination \ac{TD}. It is worth noting that the source and destination \acp{TD} have their \acp{HKID} assigned independently, thus protected with different \ac{TD} private keys.         

\nip{TEE I/O.}
A computer is comprised of various functional components. However, confidential computing has conceptually shattered the unified trust model. As a result, each component, made by different vendors, can no longer trust each other. This creates a serious impediment to efficient I/O, as untrusted devices cannot read and write data in the private memory of \acp{TEE}. To address this issue, Intel has proposed \ac{TEE} I/O in \ac{TDX} 2.0, aiming to extend the trust from a \ac{TD} to external devices. This requires changes to the devices and the \ac{TDX} platform to use a compatible protocol to establish mutual trust and enable secure communication channels. The key principle is that a \ac{TD} and a device should be able to securely exchange and verify their identities and measurements. Additionally, the data paths between a \ac{TD} and a device are not trusted and may be vulnerable to interception by attackers. Therefore, an end-to-end secure channel is necessary to protect the data transmitted between a \ac{TD} and a device. The detailed protocols for \ac{TEE} I/O can be found in the proposals~\cite{tdxio,deviceattestation}.

\section{TDX Module}
\label{sec:tdxmodule}
This section provides an in-depth analysis of the \ac{TDX} Module. We first discuss its loading process in \S\ref{sec:tdxmodule:load}, followed by an explanation of the physical and linear memory layout in \S\ref{sec:tdxmodule:address_space}. We then describe the metadata created by the \ac{TDX} Module to manage \acp{TD} in \S\ref{sec:tdxmodule:td_metadata}, and the process of context switching across security domains in \S\ref{sec:tdxmodule:context}. Additionally, we provide details about the \code{keyhole} structure (\S\ref{sec:tdxmodule:keyhole}) and memory management (\S\ref{sec:tdxmodule:mem_management}) of the \ac{TDX} Module.  

\subsection{Loading the TDX Module}
\label{sec:tdxmodule:load}
\begin{figure}[t!]
\centerline{\includegraphics[width=0.8\textwidth]{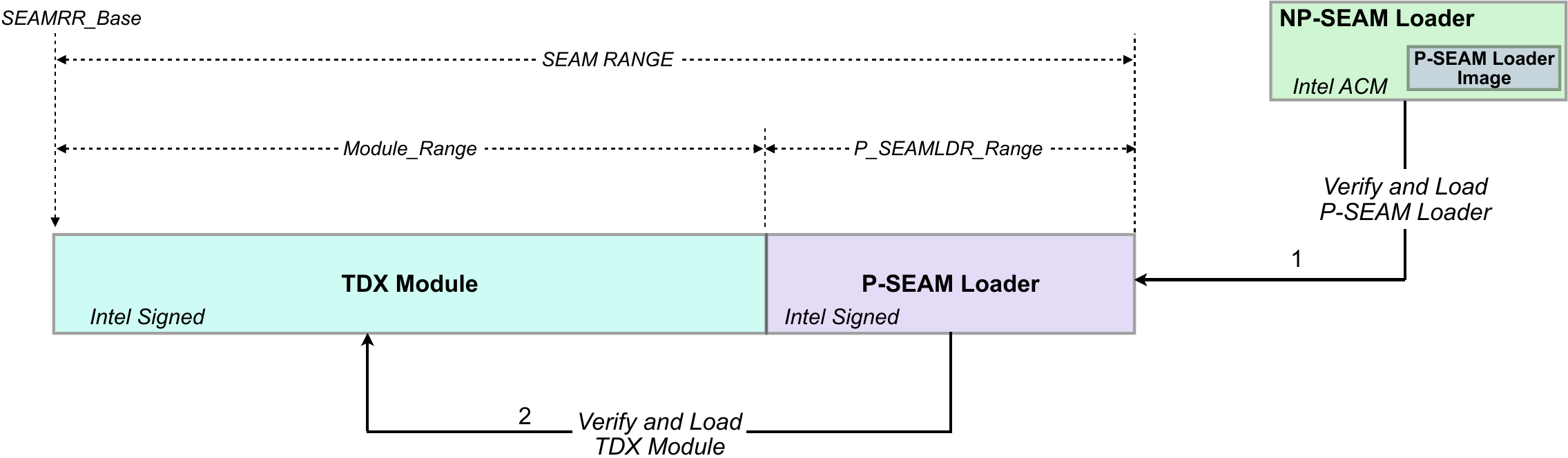}}
\caption{Loading the TDX Module}
\label{fig:tdx_module_load}
\end{figure}
\autoref{fig:tdx_module_load} illustrates the two-stage process of loading the \ac{TDX} Module. The process begins with the loading of the Intel Non-Persistent SEAM Loader (NP-SEAM Loader), which is an Intel \ac{ACM}. \acp{ACM} are Intel-signed modules that run within the internal RAM of the processor. The NP-SEAM Loader is authenticated and loaded by the Intel \ac{TXT}~\cite{txtwhitepaper} through the \code{GETSEC[ENTERACCS]} function. The NP-SEAM Loader contains the image of the Intel Persistent SEAM Loader (P-SEAM Loader), which is then verified and loaded by the NP-SEAM Loader. The P-SEAM Loader is then responsible for installing or updating the \ac{TDX} Module. 

It is important to note that both the P-SEAM Loader and the \ac{TDX} Module are loaded in the \code{\ac{SEAM} RANGE}, which is a portion of system memory reserved via UEFI/BIOS. The range's base address and size are specified by the \code{IA32\_SEAMRR\_PHYS\_BASE} and \code{IA32\_SEAMRR\_PHYS\_MASK} \acp{MSR}. This range is partitioned into \code{Module\_Range} for the \ac{TDX} Module and \code{P\_SEAMLDR\_Range} for the P-SEAM Loader. Both modules run in the \spec{\ac{SEAM} \ac{VMX} root mode} and use \code{SEAMCALL}/\code{SEAMRET} to interact with external software. The NP-SEAM Loader, P-SEAM Loader, and \ac{TDX} Module are all provided and signed by Intel, establishing a chain of trust to bootstrap the \ac{TDX} Module.

The P-SEAM Loader provides a \code{SEAMCALL} interface function \code{seamldr\_install} for loading the \ac{TDX} Module. The \ac{TDX} Module's image is pre-loaded into a memory buffer (not in the \code{\ac{SEAM} RANGE}). The physical addresses of the buffer and a \code{seam\_sigstruct} (signature of the \ac{TDX} Module) are passed as the parameters to the \code{seamldr\_install}. The \code{seam\_sigstruct} contains the hash value and the \ac{SVN} of the \ac{TDX} Module, the number of per-\ac{LP} stack pages, the number of per-\ac{LP} data pages, and the number of global data pages. These numbers are used by \code{seamldr\_install} to determine the physical/linear addresses and the sizes of the \ac{TDX} Module's various memory regions.

The \code{seamldr\_install} must be called on all \acfp{LP} serially. When \code{seamldr\_install} is called on the first \ac{LP}, an \spec{installation session} starts. On each \ac{LP}, \code{seamldr\_install} checks that the \ac{LP} is not already in an installation session (started by another \ac{LP}), and clears the \ac{LP}'s \ac{VMCS} cache. When \code{seamldr\_install} is called on the last LP, it does the following: 
\begin{enumerate}[noitemsep,topsep=0pt,partopsep=0pt]
\item 
checking the parameters to the \code{seamldr\_install},
\item 
verifying the signature of the \ac{TDX} Module,
\item
checking the \ac{SVN} of the to-be-loaded image and comparing with the resident \ac{TDX} Module,
\item \label{step:tdx_memory_constants}
determining the physical and linear addresses and sizes of the \ac{TDX} Module's various memory regions in the \code{\ac{SEAM} RANGE}: code, data, stack, page table, \code{sysinfo\_table}, \code{keyhole}, and \code{keyhole-edit} (\S\ref{sec:tdxmodule:address_space}),
\item 
mapping the regions' physical addresses to their linear addresses (\S\ref{sec:tdxmodule:address_space}),
\item 
loading the \ac{TDX} Module's binary image into the \code{\ac{SEAM} RANGE}, measuring the image, computing and verifying the \ac{TDX} Module's hash value,
\item 
setting up the \ac{TDX} Module's \code{sysinfo\_table},
\item \label{step:vmcs} 
setting up \spec{\ac{SEAM} Transfer \ac{VMCS}} on each \ac{LP} (\S\ref{sec:seam_vmcs}),
\item
recording the \ac{TDX} Module's hash, \ac{SVN}, in the P-SEAM Loader's data region.
\end{enumerate}

In addition to the \code{SEAMCALL} to install the \ac{TDX} Module, the P-SEAM Loader also provides other interface functions to shut down itself and retrieve the loader's system information. 

\subsection{Memory Layout of the TDX Module}
\label{sec:tdxmodule:address_space}
Here we discuss the physical and linear memory layout for the \ac{TDX} Module, respectively. 

\begin{figure}[ht!]
    \centering
    \includegraphics[width=0.9\textwidth]{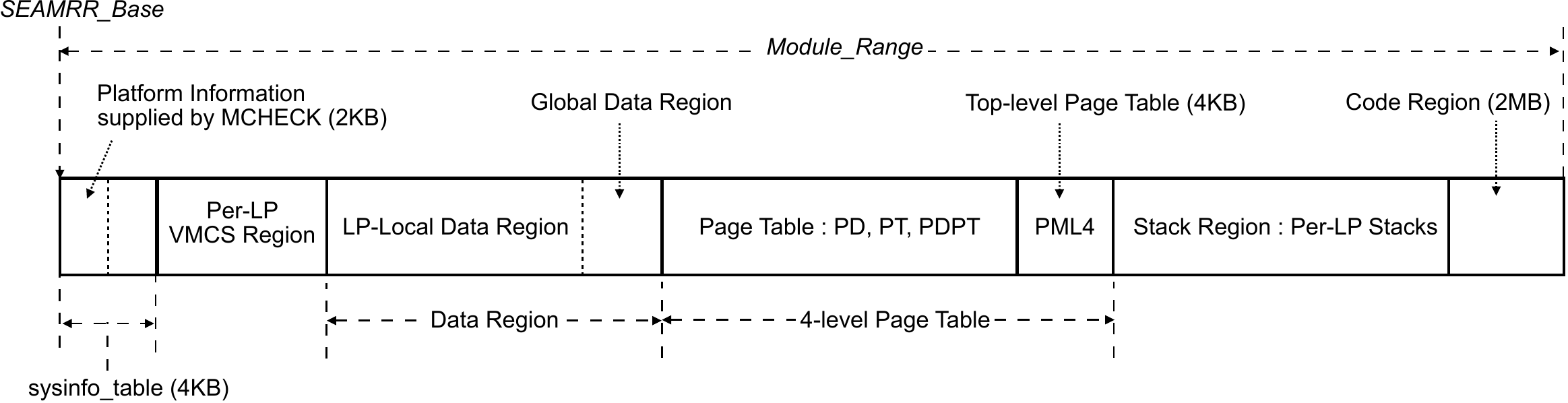}
    \caption{TDX Module Physical Memory Layout}
    \label{fig:tdx_module_pa}
\end{figure}

\nip{Physical Memory Layout.} \autoref{fig:tdx_module_pa} depicts the physical memory layout of the \ac{TDX} Module within the \code{Module\_Range}. The layout starts with a 4\,KB page that holds the \code{sysinfo\_table} of the \ac{TDX} Module. The \code{sysinfo\_table} consists of 2\,KB platform information populated by \code{MCHECK} from the NP-SEAM Loader and the next 2\,KB populated by the P-SEAM Loader with the \ac{TDX} Module's information, such as the \code{\ac{SEAM} RANGE} base address and size, the base linear addresses of the memory regions, number of \acp{LP}, and range of private \acp{HKID}. After the \code{sysinfo\_table}, there is the per-\ac{LP} \ac{VMCS} region. Each \ac{LP} has a 4\,KB \spec{\ac{SEAM} Transfer \ac{VMCS}} (see \S\ref{sec:seam_vmcs}). Following the per-\ac{LP} \ac{VMCS} region, there is the data region, which is partitioned into per-\ac{LP} data region and a global data region. Next, there is the \ac{TDX} Module's 4-level page table, followed by the per-\ac{LP} stack regions, and finally, the code region for the \ac{TDX} Module's executable code. 

\begin{figure}[ht!]
    \centerline{\includegraphics[width=1\textwidth]{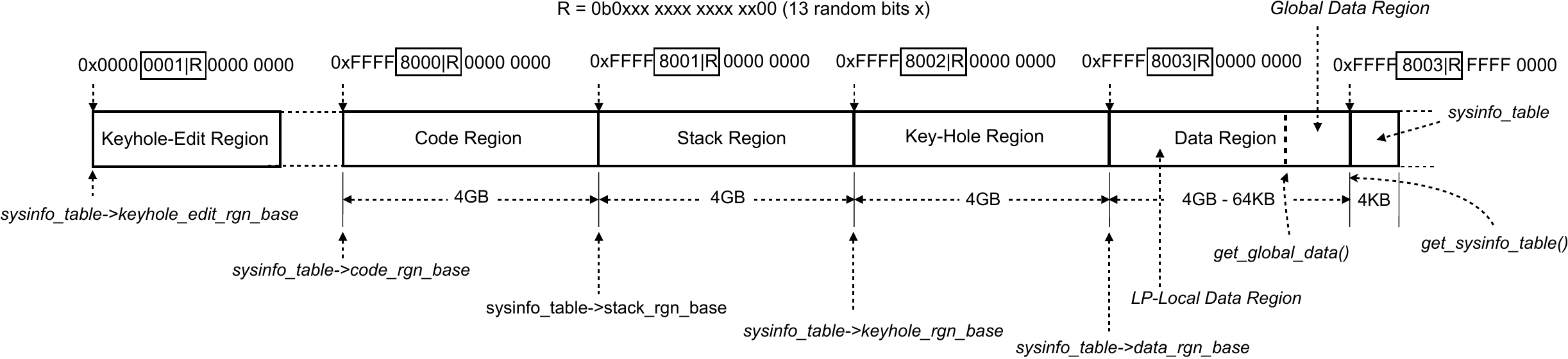}}
    \caption{TDX Module Linear Memory Layout}
    \label{fig:tdx_module_la}
\end{figure}

\nip{Linear Memory Layout.} 
The \ac{TDX} Module has its own linear address space and maintains a page table to translate addresses. \autoref{fig:tdx_module_la} illustrates the layout of the \ac{TDX} Module's linear address space, which is established by the P-SEAM Loader through the construction of the \ac{TDX} Module's page table. To prevent memory corruption attacks, the P-SEAM Loader randomizes bits $34$ to $46$ of the linear addresses, which are represented by boxes in \autoref{fig:tdx_module_la}. The linear addresses and the sizes of all regions are recorded in the fields of the \code{sysinfo\_table}. The page table entries for code, stack, data, and \code{sysinfo\_table} can be statically populated in advance and require no changes to the page table at runtime. However, the \code{keyhole} region serves to map data passed from external software dynamically during the execution of the \ac{TDX} Module. This requires the addition of the \code{keyhole-edit} region to allow runtime editing of the page table entries for the \code{keyhole}'s mapping. A detailed discussion of the \code{keyhole} and \code{keyhole-edit} regions can be found in \S\ref{sec:tdxmodule:keyhole}.

\subsection{Platform Initialization}
During the platform initialization, the hypervisor makes a \code{SEAMCALL[TDH.SYS.INIT]} to globally initialize the \ac{TDX} Module. Then, the hypervisor makes a \code{SEAMCALL[TDH.SYS.LP.INIT]} on each \ac{LP} to check and initialize per-\ac{LP} parameters, such as \code{keyholes} (\S\ref{sec:tdxmodule:keyhole}), data regions, and stack regions (\S\ref{sec:tdxmodule:address_space}). 
Next, the hypervisor allocates a global private \ac{HKID} and passes it to the \ac{TDX} Module through a \code{SEAMCALL[TDH.SYS.CONFIG]}, which also initializes the \ac{TDMR} (\S\ref{sec:tdxmodule:mem_management}).
The \code{SEAMCALL[TDH.SYS.KEY.CONFIG]} on each processor package generates a \ac{TDX} global private key and binds the key with this \ac{HKID}. This key is used to encrypt memory that holds \acf{PAMT} and \ac{TDR} (\S\ref{sec:tdxmodule:td_metadata}) of each \ac{TD}. 
Finally, the hypervisor calls \code{SEAMCALL[TDH.SYS.TDMR.INIT]} multiple times to gradually initialize the \ac{PAMT}(\S\ref{sec:tdxmodule:mem_management}) for each \ac{TDMR}. 

\subsection{Metadata for TDs}
\label{sec:tdxmodule:td_metadata}
The \ac{TDX} Module is responsible for managing the entire life-cycle of \acp{TD}. As such, it needs to maintain metadata for each \ac{TD} instance. 
The \ac{TDX} Module ensures that the memory encryption is applied to the metadata to prevent the hypervisor from accessing or modifying it. 

\begin{figure}[t!]
    \centerline{\includegraphics[width=0.8\textwidth]{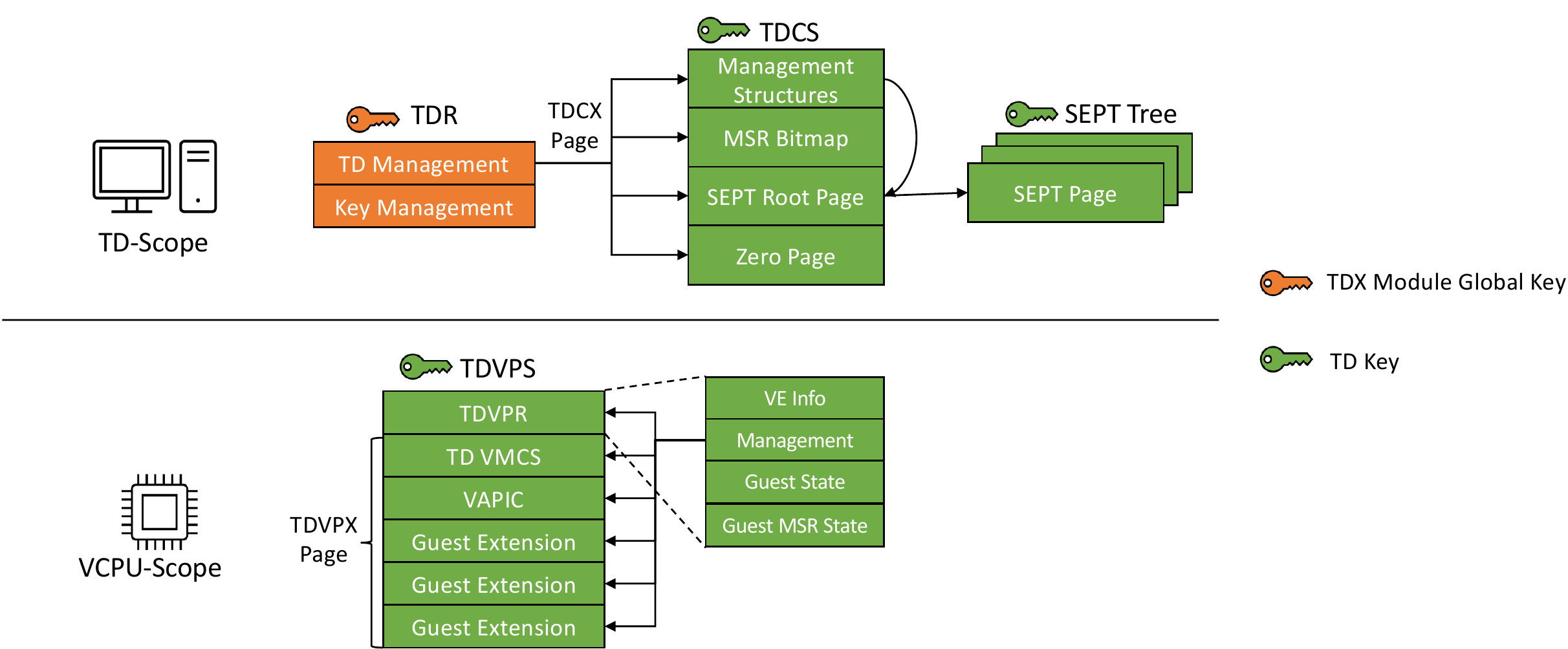}}
    \caption{Relationship of \ac{TD}'s Metadata}
    \label{fig:metadata}
\end{figure}

Each \ac{TD}'s metadata consists of the following control structures: \acf{TDR}, \acf{TDCS}, \acf{TDVPS}, and \acf{SEPT}. \autoref{fig:metadata} illustrates the relationships between these control structures.

\nip{\ac{TDR}.} \ac{TDR} is the initial structure that is created at the inception of a \ac{TD} and is destroyed when the \ac{TD} is terminated. During the entire life-cycle of the \ac{TD}, \code{SEAMCALLs} use the physical address of the corresponding \ac{TDR} to refer to the \ac{TD}. The \ac{TDR} comprises the key information for memory encryption and references to the TDCX pages (physical memory pages for the \ac{TDCS}). As the \ac{TDR} is created before the \ac{TD}'s private key is generated, it is protected with the global private key of the \ac{TDX} Module. The subsequent metadata (\ac{TDCS}, \ac{TDVPS}, and \ac{SEPT}), along with the \ac{TD}'s memory pages, can be associated with the \ac{TDR} through the \emph{owner} attribute in the \ac{PAMT} (\S\ref{sec:tdxmodule:mem_management}).

\nip{\ac{TDCS}.} \ac{TDCS} is a control structure that manages the operations and stores the state at the scope of a \ac{TD}. It is comprised of four continuous TDCX memory pages, each allocated for a specific purpose, such as \ac{TD}'s management structures, \ac{MSR} bitmaps, \ac{SEPT} root page, and a special zero page. \ac{TDCS} is encrypted with the \ac{TD}'s private key, which is generated when the \ac{TDR} is created.

\nip{\ac{TDVPS}.} \ac{TDVPS} is a control structure for each virtual CPU of a \ac{TD}. It is comprised of six memory pages, starting from a TDVPR page that contains references to multiple TDVPX pages. The first TDVPR page holds the fields for \ac{VE} information, virtual CPU management, guest state, and guest \ac{MSR} state. The second page is for the \spec{\ac{TD} Transfer \ac{VMCS}} (\S\ref{sec:td_vmcs}), which controls the \ac{TD}'s entry and exit. The third page is a virtual APIC (VAPIC) page, followed by three pages for guest extension information. Like the \ac{TDCS}, the \ac{TDVPS} is also protected by the \ac{TD}'s private key.

\nip{\ac{SEPT}.} For legacy \acp{VM}, hypervisors manage address translations from \ac{GPA} to \ac{HPA} using \ac{EPT}. However, in \ac{TDX}, guest address translations must be protected from untrusted hypervisors. To achieve this, \ac{TDX} has two types of \ac{EPT}: \spec{\acf{SEPT}} and \spec{Shared \ac{EPT}}. \spec{\acl{SEPT}} is used to translate addresses of a \ac{TD}'s private memory and is protected by the \ac{TD}'s private key. The reference to the \ac{SEPT} and the \ac{SEPT} root page are stored in the \ac{TDCS}. \spec{Shared \ac{EPT}}, on the other hand, is used to translate addresses for memory explicitly shared by the \ac{TD} with a hypervisor, such as in the case of virtualized I/O. It remains under the control of the hypervisor. The guest kernel in the \ac{TD} can determine which memory pages to share by setting the \spec{shared} bit in the \ac{GPA}. Shared memory pages are not encrypted with the \ac{TD}'s private key.

\subsection{Context Switches}
\label{sec:tdxmodule:context}
There are two types of context switches for \ac{TDX}: the first occurs between the hypervisor and the \ac{TDX} Module, while the second occurs between \acp{TD} and the \ac{TDX} Module. We delve into each of these in more detail.

\nip{Hypervisor $\leftrightarrow$ \ac{TDX} Module.}\label{sec:seam_vmcs}
In \ac{TDX}, a hypervisor is prohibited from directly managing \acp{TD}. Instead, it must interact with the \ac{TDX} Module through \code{SEAMCALL} interface functions. When a \code{SEAMCALL} is made, the processor transitions from the \spec{\ac{VMX} root mode} to the \spec{\ac{SEAM} \ac{VMX} root mode}. The \ac{TDX} Module's \spec{\ac{SEAM} Transfer \ac{VMCS}} is loaded. This \spec{\ac{SEAM} Transfer \ac{VMCS}} is set up by the P-SEAM Loader for each \ac{LP} and is stored within the \code{Module\_Range}.

The repurposing of \ac{VMCS} for context switches between the hypervisor and the \ac{TDX} Module may seem confusing initially, as the hypervisor is not a ``guest \ac{VM}'' and the \ac{TDX} Module is not a ``host hypervisor.'' We can disregard the guest/host concept and only view the \spec{\ac{SEAM} Transfer \ac{VMCS}} as a means of switching the execution context between the hypervisor and the \ac{TDX} Module. 

\begin{figure}[t!]
    \centerline{\includegraphics[width=0.7\textwidth]{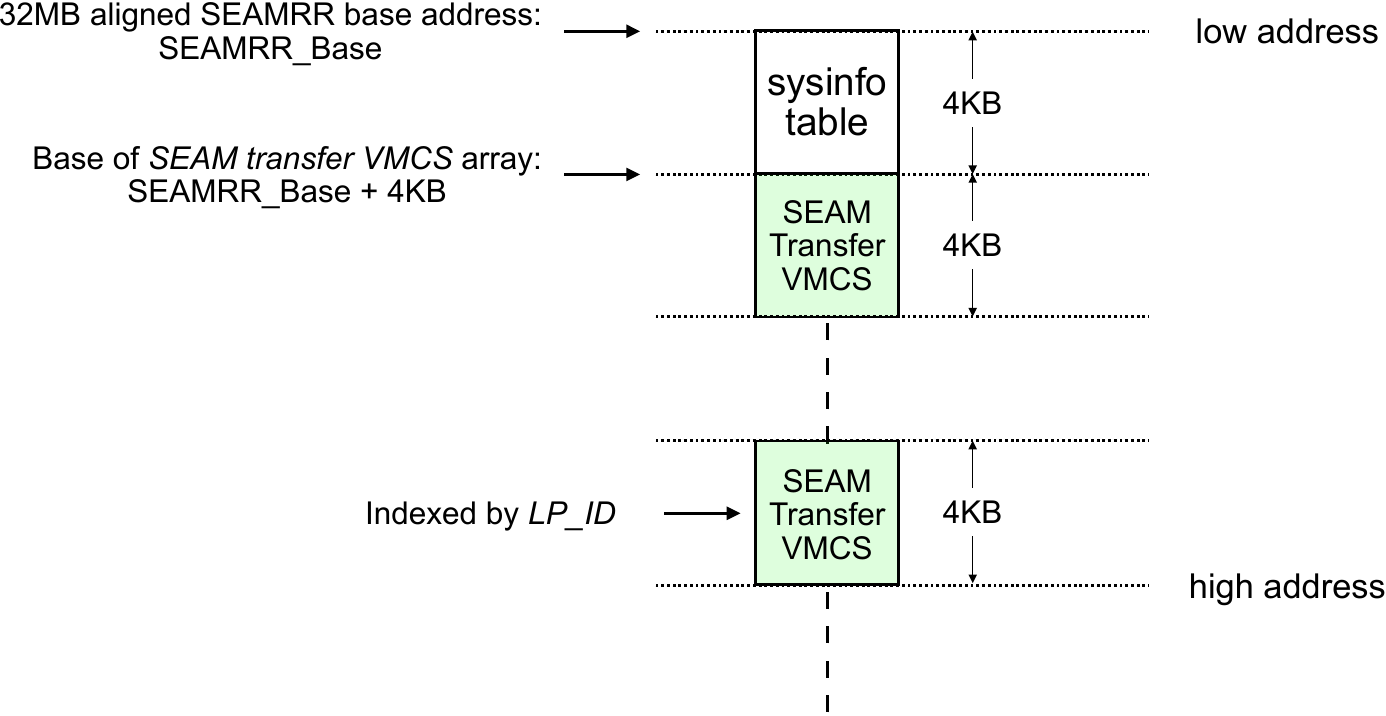}}
    \caption{Per-LP \spec{\ac{SEAM} Transfer \ac{VMCS}} Layout}
    \label{fig:seam_vmcs}
\end{figure}
\autoref{fig:seam_vmcs} depicts the location and layout of the \spec{\ac{SEAM} Transfer \ac{VMCS}} regions in the \code{Module\_Range}, which begins at the \code{SEAMRR\_Base}. The first 4\,KB page in the \code{Module\_Range} is the \code{sysinfo\_table}. Starting from \code{SEAMRR\_Base + 4\,KB}, there is an array of per-\ac{LP} \spec{\ac{SEAM} Transfer \ac{VMCS}} regions. Each region is a 4\,KB page. This array is indexed by the identifier of the \ac{LP}, denoted by \code{LP\_ID}. 

When a \code{SEAMCALL} instruction is executed, the processor searches for the \ac{VMCS} address based on the current \code{LP\_ID}. The address is determined by: \code{SEAMRR\_Base + 4\,KB + (\ac{LP}\_ID $\times$ 4\,KB)}. The \ac{TDX} Module's state is stored in the \emph{host-state area} of the \ac{VMCS}. For instance, the host \code{RIP} is set to the \code{tdx\_seamcall\_entry\_point} of the \ac{TDX} Module, and the host \code{CR3} is set to the physical address of the \ac{TDX} Module's \code{PML4} base. Moreover, the host \code{FS\_BASE} is set to the linear address of the \code{sysinfo\_table} and the host \code{GS\_BASE} is set to the per-LP data region. 

When the \ac{LP} transitions into the \ac{TDX} Module through the \code{SEAMCALL} instruction, the information stored in the \spec{\ac{SEAM} Transfer \ac{VMCS}} is loaded onto the processor. Therefore, the \code{FS\_BASE} and \code{GS\_BASE} now point to the \code{sysinfo\_table} and the local data region of the \ac{TDX} Module, respectively. The \code{CR3} register points to the \ac{TDX} Module's page table, thereby switching \ac{MMU} to operate in the \ac{TDX} Module's linear address space. The \ac{TDX} Module starts to handle the \code{SEAMCALL} and dispatch it to corresponding interface functions. 

\nip{TD $\leftrightarrow$ \ac{TDX} Module.}
\label{sec:td_vmcs}
In traditional virtualization, the hypervisor handles \ac{VM} exits, which are controlled by the \ac{VM}'s \spec{Transfer \ac{VMCS}}. Each \ac{VMCS} is associated with one virtual CPU and stores the virtual CPU state for recovering the guest execution in the next \ac{VM} resume. However, this operation leaks the virtual CPU state as \ac{VMCS} is visible to hypervisors. In \ac{TDX}, synchronous \code{TDCALLs} or asynchronous \ac{TD} exits are designed to trap into the \ac{TDX} Module. This is controlled by the \spec{\ac{TD} Transfer \ac{VMCS}}, which is set up when a virtual CPU of a \ac{TD} is created and is stored in the \ac{TD}'s \ac{TDVPS}. The \ac{TDVPS} is encrypted with the \ac{TD}'s private key (\S\ref{sec:tdxmodule:td_metadata}). Therefore, the \spec{\ac{TD} Transfer \ac{VMCS}} is inaccessible to untrusted hypervisors. When a \ac{TD} calls a \code{TDCALL} or triggers a \ac{TD} exit, the \ac{LP} loads the state of the \ac{TDX} Module stored in the \spec{\ac{TD} Transfer \ac{VMCS}} to switch context.  

In \ac{TDX}, certain \ac{TD} exits cannot be fully handled by the \ac{TDX} Module and instead require a hypervisor to emulate certain operations, such as port I/O, \code{HLT}, \code{CPUID}, and more. However, traditional hypervisors have access to the entire virtual CPU states and memory, exposing more information than necessary to handle these exits. \ac{TDX} addresses this issue by introducing a new mechanism for handling TD exits. All \ac{TD} exits first trap into the \ac{TDX} Module, which injects a \acf{VE} into the \ac{TD} to handle the exit. The \ac{TD}'s guest kernel includes a corresponding \ac{VE} handler that prepares a minimized set of parameters and invokes a \code{TDCALL} to re-enter the \ac{TDX} Module. At this point, the \ac{TDX} Module can safely ask the hypervisor to handle the requests with minimal exposure of sensitive information.

\subsection{Keyholes}
\label{sec:tdxmodule:keyhole}
All memory buffers passed through \code{SEAMCALLs} use their physical addresses as references. The \ac{TDX} Module must map these buffers into its own linear address space to access them. This mapping process is facilitated by the \code{keyhole} and \code{keyhole-edit} regions, which serve as temporary ``leases'' of linear addresses.

The \code{keyhole} region is a reserved linear address range specifically for address mapping. The region is comprised of an array of \code{keyholes}. This array is further divided into $128$-\code{keyhole} segments, with each segment assigned to one \ac{LP}. The \ac{TDX} Module organizes free \code{keyholes} in an LRU list when setting up per-\ac{LP} data structures. Each \code{keyhole} corresponds to a 4\,KB aligned linear address and links to a physical memory page. Since multiple memory buffers can exist within the same memory page, each \code{keyhole} maintains a \emph{reference count} to track the number of referenced buffers in the page. 

When the \ac{TDX} Module is installed by the P-SEAM Loader, all the linear addresses of \code{keyholes} are mapped to an empty physical address. This is achieved by setting all the leaf-level page table entries (PTEs) for the \code{keyhole} region in the \ac{TDX} Module's page table to zero. Simultaneously, the physical addresses of the corresponding PTEs for the \code{keyholes} are mapped to the \code{keyhole-edit} region. This enables the \ac{TDX} Module to locate and modify the \code{keyhole}'s address mappings in its page table during runtime.

When processing a \code{SEAMCALL} that refers to an external memory buffer with a physical address, the \ac{TDX} Module checks if the buffer's memory page is already mapped by a \code{keyhole}. If so, it increments the \code{keyhole}'s \spec{reference count} and returns the mapped linear address. If not, it selects a free \code{keyhole} from the LRU list and maps the linear address of this \code{keyhole} to the page table by updating the corresponding PTE referenced in the \code{keyhole-edit} region. Once the buffer is mapped, the \ac{TDX} Module can access it using the \code{keyhole}'s linear address. At the end of each \code{SEAMCALL}, the \spec{reference counts} of corresponding \code{keyholes} decrement, and any non-referenced \code{keyholes} return to the LRU list. 

\subsection{Physical Memory Management}
\label{sec:tdxmodule:mem_management}
The \ac{TDX} Module manages physical memory by using a set of \acfp{TDMR} and their control structures, \acfp{PAMT}. \acp{TDMR} are constructed by the hypervisor based on a list of \acp{CMR}, which are the memory regions that can be used for \ac{TD}'s private memory or metadata. These regions are subject to \ac{MKTME} encryption and \ac{TDX} memory integrity protection. This list of \acp{CMR} is prepared by the UEFI/BIOS.

Each \ac{TDMR} is a single range of physical memory that is 1\,GB aligned and has a size that is an integral multiple of 1\,GB, but does not necessarily need to be a power of two. Two \acp{TDMR} cannot overlap. A \ac{TDMR} may contain \spec{reserved areas} that cannot be used by the \ac{TDX} Module. A reserved area is an array of 4\,KB aligned memory pages (each page is 4\,KB). Memory in a \ac{TDMR}, except for the reserved areas, must be convertible. It should be noted that \ac{TDMR} configuration is managed by software without using hardware range registers. 

The \ac{TDX} Module uses \ac{PAMT} to track page attributes of each physical memory page in a \ac{TDMR}. The attributes contain the information about the \spec{page owner}, \spec{page type}, and \spec{page size}. The page attributes allow the \ac{TDX} Module to ensure that a physical memory page in a \ac{TDMR} has a proper type and is only assigned to at most one \ac{TD}. When a page is assigned to a \ac{TD}'s private memory, the \ac{TDX} Module can check whether the page size in the \ac{SEPT} and \ac{PAMT} are consistent.

A \ac{PAMT} is divided into blocks, where each block tracks page addresses within the 1\,GB size range. Each block has three levels to track metadata for pages with sizes 4\,KB, 2\,MB, and 1\,GB, respectively. The first level tracks a single 1\,GB page, the second level tracks $512$ 2\,MB pages, and the third level tracks $512 \times 512$ 4\,KB pages. Given a physical address, the \ac{TDX} Module can perform a \ac{PAMT} hierarchical walk to retrieve its page attributes for a sanity check. 

The \ac{TDX} Module manages the data structure by updating the attributes of each page it uses during runtime. Any operation that requires accessing, removing, or adding a page causes the \ac{TDX} Module to walk through \ac{PAMT} to adjust the page attributes and check corresponding access rights. The memory for \ac{PAMT} is allocated by the hypervisor and is encrypted with the \ac{TDX} Module's global private key.
\section{Memory Protection}
\label{sec:memprotect}
A \ac{TD}'s memory is divided into \spec{private memory} and \spec{shared memory}. The private memory is only accessible by the \ac{TD} and the \ac{TDX} Module. The shared memory is also accessible by the hypervisor and is used for operations that require the cooperation from the hypervisor, such as networking, I/O, and \ac{DMA}. \ac{TDX} protects the confidentiality and integrity of a \ac{TD}'s private memory. 

\subsection{HKID Space Partitioning}
\label{sec:memprotect:MKTME}
\begin{figure}[ht!]
\centerline{\includegraphics[width=0.8\textwidth]{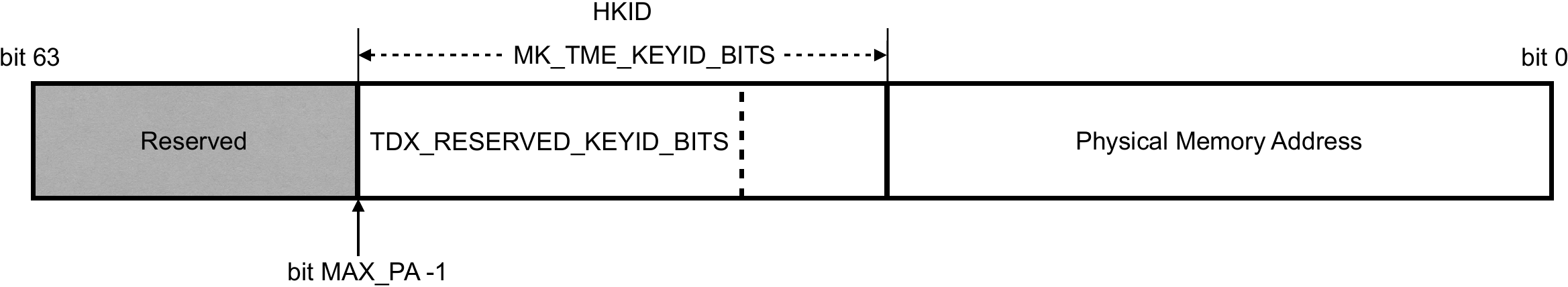}}
\caption{HKID Layout in Physical Memory Address}
\label{fig:hkid}
\end{figure}

The \ac{HKID} space is partitioned once during the boot process into two ranges, \spec{private \acp{HKID}} and \spec{shared \acp{HKID}}. Only software in the \ac{SEAM} mode, namely the \ac{TDX} Module and \acp{TD}, can read and write memory whose contents are encrypted by keys associated with private \acp{HKID}. Keys associated with shared \acp{HKID} can be used to encrypt memory outside the SEAM mode, such as the memory of legacy VMs and the host kernel.

When the hypervisor requests the \ac{TDX} Module to establish a \ac{TD}, it allocates a private \ac{HKID} for the \ac{TD}. The \ac{TDX} Module, using the \code{PCONFIG} instruction, asks \ac{MKTME} to generate an unique random key for the \ac{HKID}. This key is called the \ac{TD}'s \spec{ephemeral private key}. It is used to encrypt all the private memory and metadata of the \ac{TD} and is never exposed outside \ac{MKTME}. This \code{$\langle\ac{HKID}, key\rangle$}  binding is valid for the lifetime of the \ac{TD}.

A physical memory page associated with a HKID stores the HKID in the upper bits of the page's physical address, as shown in \autoref{fig:hkid}. At boot time, the number of bits used for \acp{HKID} (\code{MK\_TME\_KEYID\_BITS}) and the number of bits used for private \acp{HKID} (\code{TDX\_RESERVED\_KEYID\_BITS}) are set in the \code{IA32\_TME\_ACTIVATE} \ac{MSR}. The \code{IA32\_MKTME\_KEYID\_PARTITIONING} \ac{MSR} can be used for reading the numbers of private and shared \acp{HKID}. Intel reserves a range of upper bits in the 64-bit physical address. The \ac{HKID} uses these reserved bits. The remaining bits following the \ac{HKID} correspond to the physical memory address. The upper bits of the \ac{HKID} field, the \code{TDX\_RESERVED\_KEYID\_BITS} are reserved for private \acp{HKID}. For example, if \code{MK\_TME\_KEYID\_BITS} is $6$ and \code{TDX\_RESERVED\_KEYID\_BITS} is $4$, then \acp{HKID} from $0$ to $3$ are \spec{shared}, and \acp{HKID} from $4$ to $63$ are \spec{private}.

The hypervisor and the \ac{TDX} Module configure the memory encryption by setting the \ac{HKID} in the upper bits of the physical address of a memory page. The hypervisor can only use shared \acp{HKID}, while the \ac{TDX} Module can use both shared and private \acp{HKID}. An exception will be raised if any software executing outside \ac{SEAM} mode tries to access memory through a physical address with a private \ac{HKID}.

\subsection{TD Memory Integrity Protection}
\label{sec:memprotect:TD_memory_integrity}
\ac{TDX} always protects the integrity of the \ac{TD}'s private memory content. 
This protection is required because an entity outside the \ac{SEAM} mode,\eg a malicious hypervisor or a \ac{DMA} device, can write to the \ac{TD}'s private memory. \ac{TDX} cannot prevent such modification, but it can detect and flag it. It prevents a \ac{TD} or the \ac{TDX} Module from reading or executing the modified content. To detect such modifications, \ac{TDX} supports two memory integrity modes that can be configured on a system:
\begin{enumerate}[noitemsep,topsep=0pt,partopsep=0pt]
    \item{\acf{Li}}: memory integrity is protected by a \spec{\ac{TD} Owner bit}.
    \item{\acf{Ci}}: memory integrity is protected by a \ac{MAC} and a \spec{\ac{TD} Owner bit}.
\end{enumerate}
Both \ac{Li} and \ac{Ci} apply to a physical memory segment with the size of a cache line and whose address is cache line aligned.
\ac{Ci} can detect modifications made by direct physical access to the memory or bit flips, such as the Rowhammer attack~\cite{rowhammer-ieee-2014}, which \ac{Li} cannot detect.

In addition to \ac{Li} and \ac{Ci}, if a program outside the \ac{SEAM} mode reads the private memory of a \ac{TD} or the \ac{TDX} Module, the read will always return zeros. This is to prevent ciphertext cryptanalysis and side channels in which a program outside the \ac{SEAM} mode could determine whether a program in the \ac{SEAM} mode change the memory content. 

If a \ac{TD} or the \ac{TDX} Module writes to a memory segment belonging to a \ac{TD}'s private memory, the corresponding \spec{\ac{TD} Owner bit} is set to $1$. Due to the way a \ac{TD}'s memory is set up, all \spec{\ac{TD} Owner bits} of a \ac{TD}'s private memory should be set to $1$. However, if an entity outside the \ac{SEAM} mode writes to a segment belonging to the private memory, the corresponding \spec{\ac{TD} Owner bit} is cleared to $0$. Later, when the \ac{TD} or the \ac{TDX} Module reads the segment, the segment is marked as \emph{poisoned}. If the reader is the \ac{TD}, this \spec{poisoned} marking causes a \ac{TD} exit for the \ac{TD}. The \ac{TDX} Module can capture this \ac{TD} exit and put the \ac{TD} into a \spec{fatal} state, which prevents any further entry into the \ac{TD} and leads to the tearing down of the \ac{TD}. If the \ac{TDX} Module reads the \spec{poisoned} content, the \ac{TDX} Module and the \ac{TDX}'s hardware extension in the processor are marked as \spec{disabled}. Any further \code{SEAMCALLs} leads to the \code{VMFailInvalid} error. 

If \ac{Ci} is enabled, the processor generates a $128$-bit MAC key during system initialization. On each write, \ac{TDX} uses this key to calculate and store a 28-bit \ac{MAC} in the \ac{ECC} memory corresponding to the cache line. On each read, the memory controller recalculates the \ac{MAC} and compares it with the value read from the \ac{ECC} memory. The mismatch indicates integrity or authenticity violation and results in the cache line being marked as \spec{poisoned}. The \ac{MAC} is calculated over:
(1) the ciphertext (encrypted content of the cache line), 
(2) the tweak values used for AES-XTS encryption, 
(3) the \spec{\ac{TD} Owner bit}, and 
(4) the 128-bit MAC key.

\section{Remote Attestation}
\label{sec:remote_attestation}
The attestation of a \ac{TD} consists of generating a local attestation report, which can be verified on the platform and then extending this report with digital signatures and certificates to enable remote attestation of the \ac{TD} off the platform. We first describe the overall process of generating and extending a local \ac{TD} \spec{report} in \S\ref{ssec:attestation_process}. Then we review the setup and the configuration of the host \ac{TDX} platform to enable remote attestation in \S\ref{ssec:setup}.  Finally, we provide details on using remote attestation for establishing a secure channel and encrypted boot in \S\ref{ssec:secure_channel}.

\subsection{Attestation Process}
\label{ssec:attestation_process}
Several steps are involved when generating and extending a local \spec{report} of a \ac{TD} to enable remote attestation. The first step is to take measurements of the loaded software during the build-time and runtime of the \ac{TD}. The next step is to retrieve the \ac{TD}'s measurements and platform \ac{TCB} information, \ie generating a \ac{TD} \spec{report}. The final step is to derive a \spec{quote} from the \ac{TD} \spec{report}. A third party can use the \spec{quote} to verify whether the \ac{TD} runs on a genuine \ac{TDX} platform with the expected \ac{TCB} versions and software measurements.

\nip{Taking Measurements.} \ac{TDX} provides two types of measurement registers for each TD: a build-time measurement register called \ac{MRTD} and four \acp{RTMR}. These measurement registers are comparable to the \ac{TPM}'s \acp{PCR}, see Table~\ref{tdx-tpm-table} derived from~\cite{tdxvirtualfw} showing the mapping between \ac{TDX} measurement registers and \ac{TPM} \acp{PCR}.

\begin{table}[t!]
\caption{Mapping of \ac{TDX} Measurement Registers and \ac{TPM} PCRs}
\label{tdx-tpm-table}
\small
\begin{tabular}{l|l|l} 
 \hline
 \textbf{\ac{TDX} Measurement Registers}  & \textbf{\ac{TPM} \acp{PCR}} & \textbf{Usage} \\ 
 \hline
 \ac{MRTD}   & \ac{PCR}[0]  & Virtual firmware  \\
 \hline
 \ac{RTMR} [0] & \ac{PCR}[1,7] & Virtual firmware data + configuration \\
 \hline
 \ac{RTMR} [1] & \ac{PCR}[2-5] & OS kernel + \spec{initrd} + boot parameters \\ 
 \hline
 \ac{RTMR} [2] & \ac{PCR}[8-15] & OS application \\
 \hline
 \ac{RTMR} [3] & N/A & Reserved \\
 \hline
\end{tabular}
\end{table}

The \ac{MRTD} contains a measurement of the \ac{TD} build process.  At the \ac{TD} creation, when the hypervisor adds initial memory pages to the \ac{TD}, it extends the \ac{MRTD} in the \ac{TDCS} with measurements of these pages. The hypervisor calls \code{SEAMCALL[TDH.MEM.PAGE.ADD]}  to add a page to the TD's memory and to initiate the measurement of the page. It first calculates a \spec{SHA384} update over the ASCII string ``MEM.PAGE.ADD'' and the \ac{GPA} of the page. Then it extends the \ac{MRTD} with the hash value. Once the page is copied into the \ac{TD}’s memory, \ie mapped and available in the \ac{SEPT}, the hypervisor calls \code{SEAMCALL[TDH.MR.EXTEND]} multiple times to measure the content of the page. The page is measured in blocks of 256\,B. For each block, the extension operation first calculates a \spec{SHA384} update over the ASCII string ``MR.EXTEND'' and the \ac{GPA} of the block. Second, it calculates another \spec{SHA384} update over the content of this block. Both hash values extend the \ac{MRTD}. These initial pages contain the \ac{TD}'s virtual firmware. The \ac{MRTD}'s measurement does not include pages containing control structures, \ie \ac{TDR}, \ac{TDCS}, and \ac{TDVPS}, nor the \ac{SEPT}. After the initial set of pages are added, the hypervisor finalizes the \ac{MRTD} measurement using the \code{SEAMCALL[TDH.MR.FINALIZE]}. This disables future operations to extend the \ac{MRTD}. For example, when initializing a \ac{TD}, KVM, as a Linux hypervisor, measures the TDVF~\cite{tdvf} (virtual firmware of a \ac{TD}) code into the \ac{MRTD}. 

\acp{RTMR} are general measurement registers labeled $0$ through $3$ for \ac{TD}'s runtime measurements. A \ac{TD} can use these registers to provide a measured boot, \ie measuring all software loaded after booting. These measurement registers are initialized to zero. The TD calls \code{TDCALL[TDG.MR.RTMR.EXTEND]} to extend the content of a \ac{RTMR}. The arguments of this call consist of an index to the measurement register and an 64\,B-aligned physical address of the 48\,B extension buffer containing the value. This call calculates a \spec{SHA384} hash over the current value of the given index measurement register concatenated with the value in the extension buffer, as follows \code{\ac{RTMR}[index] = SHA384(\ac{RTMR}[index] || value)}. For example, TDVF measures the static/dynamic configuration data into the \code{\ac{RTMR}[0]} and the OS kernel, boot parameters, and \spec{initrd} into the \code{\ac{RTMR}[1]}. 

\nip{Generating a TD Report.} A TD \spec{report} is generated inside a TD. The TD calls \code{TDCALL[TDG.MR.REPORT]}, which is the \ac{TDX} Module's report function, with a newly initialized report structure and some user report data, named \code{REPORTDATA}. The \code{REPORTDATA} is 64\,B and it can be used as a \spec{nonce} to verify freshness of the TD \spec{report}. To service the call, the \ac{TDX} Module invokes the newly added \code{SEAMOPS[SEAMRERPORT]} instruction with the \ac{TD}'s measurements and \code{REPORTDATA}. The CPU adds \ac{TCB} information related to the \ac{SEAM} and returns a \ac{TD} \spec{report}. This \ac{TD} \spec{report} is integrity protected using an HMAC key maintained by the CPU. The HMAC key is only available to the CPU. The \ac{TDX} Module returns this \spec{report} to the TD. Using the \code{EVERIFYREPORT2} instruction, an enclave can verify the \spec{report} on the same platform but not off the platform.

\begin{figure}[t!]
\centerline{\includegraphics[width=0.7\textwidth]{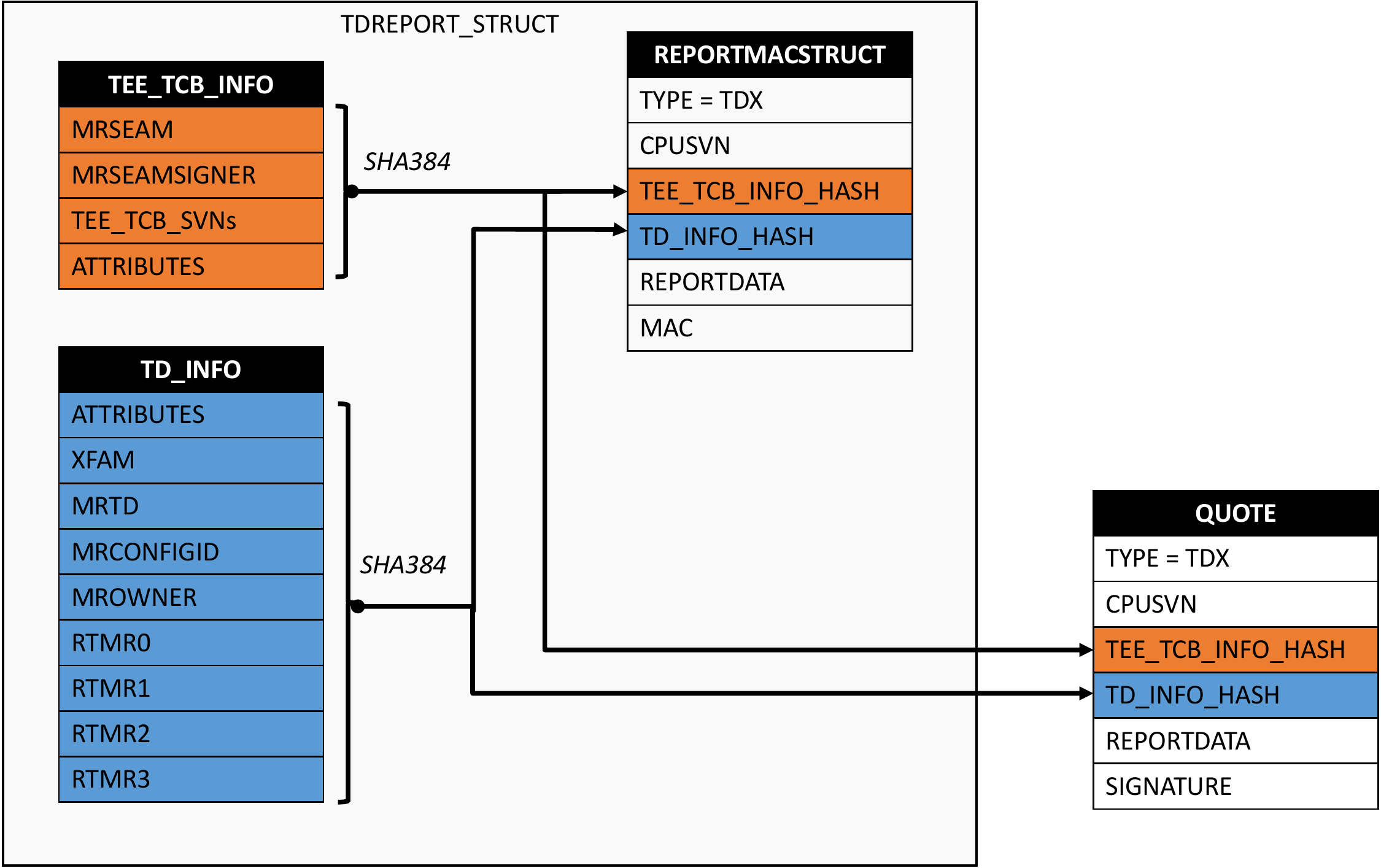}}
\caption{TD Report Structure}
\label{fig:td_report}
\end{figure}

\autoref{fig:td_report} illustrates the \ac{TD} \spec{report} consisting of three components: \code{REPORTMACSTRUCT}, \code{TEE\_TCB\_INFO}, and \code{TD\_INFO}. The \code{REPORTMACSTRUCT} structure contains header information specifying the structure type as \ac{TDX} and has fields for CPU \ac{SVN} and hashes of the \code{TEE\_TCB\_INFO} and \code{TD\_INFO} components. It also includes the \code{REPORTDATA} provided as the input to the \code{TDCALL[TDG.MR.REPORT]} function.  Finally, there is an HMAC over the entire header that protects \code{TEE\_TCB\_INFO} and \code{TD\_INFO} components. The \code{TEE\_TCB\_INFO} structure contains information about the \ac{SVN} and measurements of the \ac{TDX} Module. The \code{TD\_INFO} contains TD attestable properties. Examples of these properties include the initial \ac{TD} configuration and values of the measurement registers. 

\nip{Deriving a Quote.} To enable verification off the platform by a third party, the \ac{TD} \spec{report} must be converted into a \spec{quote}. \ac{TDX} tends to reuse the remote attestation mechanism of \ac{SGX}. 
A TD makes a call to request the \acf{QE} running on the host platform to sign the \ac{TD} \spec{report}. This call can be implemented over a \spec{VSOCK} or a \code{TDCALL[TDG.VP.VMCALL]}, depending on how the quoting service is provided on the platform. The \ac{QE} calls the \code{EVERIFYREPORT2} instruction to verify the TD \spec{report}'s HMAC. If this call is successful, the \ac{QE} signs the TD \spec{report} using its certified attestation key to generate a \spec{quote}. This operation basically replaces the MAC integrity protection of the TD \spec{report} with the digital signature protection, allowing any party to verify the provenance and integrity of the \spec{quote} using public key certificates. Section~\S\ref{ssec:setup} describes the operations for enabling the local attestation infrastructure on the platform to support remote attestation.

\subsection{Platform Setup}
\label{ssec:setup}
Configuring the attestation infrastructure involves registering the platform with the Intel \ac{PCS}, running architectural enclaves for generating \spec{quotes}, and retrieving certificates required for verifying \spec{quotes}. Intel extends the existing \ac{DCAP}~\cite{dcapwhitepaper} to support remote attestation for \ac{TDX}.

\nip{Registration.}
On multiple-package platforms, platform keys are derived at platform assembly time. 
These keys are shared between CPU-packages and are encrypted by the CPU's unique hardware key. \acfp{PCK} are derived from the platform keys and used for certifying (signing) attestation keys. Since \acp{PCK} are not recognized by the attestation infrastructure, they must be registered with Intel \ac{PCS}.

To register a platform, we need to run the \spec{\ac{PCK} Cert ID Retrieval Tool} to extract a manifest from the platform. This manifest contains information on CPU packages, \eg CPU ID (128-bit), \ac{SVN}, and hardware \ac{TCB} information. When the Intel \ac{PCS} gets the register server request, it checks whether CPUs and \ac{TCB} are in good standing before issuing a \ac{PCK} certificate. The manifest is signed with keys derived from the CPU package’s hardware keys and the Intel \ac{PCS} checks whether these signatures are valid.  If registration succeeds, the Intel \ac{PCS} returns an Intel-issued certificate for the \ac{PCK}.

Typically in \ac{DCAP}, a \ac{PCCS} runs on the host platform to facilitate \ac{PCK} certificate retrieval. This service can run anywhere. It forwards the \ac{PCK} requests from the \spec{\ac{PCK} Cert ID Retrieval Tool} to the Intel \ac{PCS} and caches the returned \ac{PCK} certificates locally. The Intel \ac{PCS} also provides certificates and revocation lists for \acp{PCK} in all genuine Intel platforms. \ac{PCCS} maintains local caches of these artifacts as well. 

Before registering, a platform must have the appropriate UEFI/BIOS settings and access to the Intel \ac{PCS}. Both \ac{TDX} and \ac{SGX} must be enabled in the UEFI/BIOS on the host platform.  An Intel account is required for retrieving API keys for registering a platform with the Intel \ac{PCS}. If the \ac{PCCS} is utilized, it must be configured with the API keys and Intel \ac{PCS} server's address.

\begin{figure}[t!]
\centerline{\includegraphics[width=0.50\textwidth]{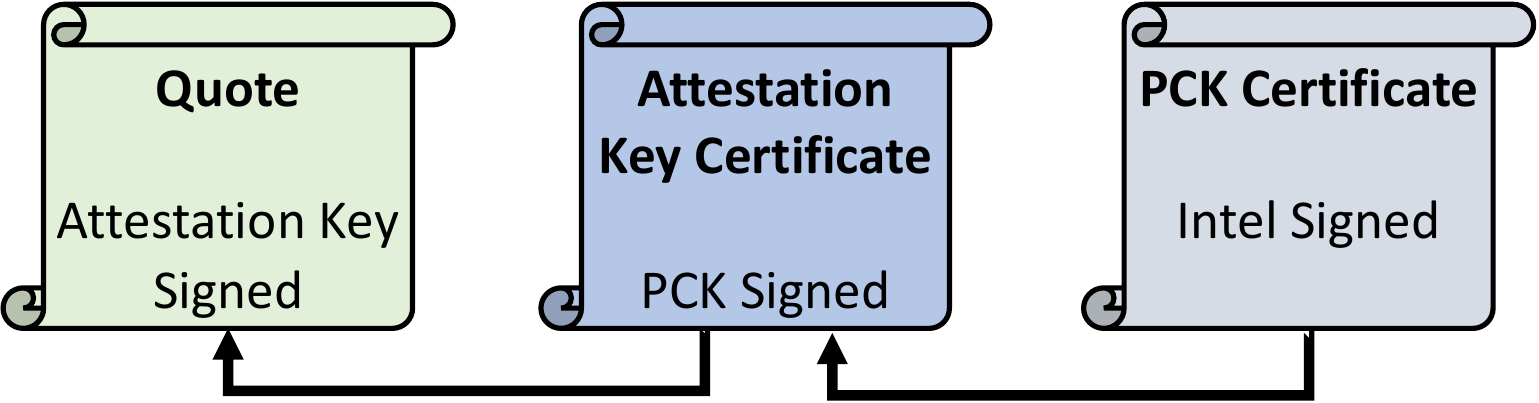}}
\caption{Quote Certificate Chain}
\label{fig:quote_chain}
\end{figure}
\nip{Architectural Enclaves.}
To enable \spec{quote} generation on the platform. Intel provides two architectural enclaves: \ac{PCE} and \acf{QE}. The \ac{PCE} acts as a local certification authority for the \ac{QE}. In its initialization process, the \ac{QE} generates an attestation key pair. It sends the public part to the \ac{PCE}. The \ac{PCE} authenticates that this is a legitimate \ac{QE} on the platform and then signs the attestation public key certificate with the \ac{PCK}. This signature creates a \spec{quote} certificate chain from an Intel-issued \ac{PCK} certificate to the \ac{QE} attestation public key. \autoref{fig:quote_chain} illustrates the \spec{quote}'s certificate chain. The \ac{PCK} certificate is used for verifying the \ac{QE} attestation public key certificate, and the \ac{QE} attestation public key in turn for verifying the signature on the \spec{quote}. 	

\begin{figure}[t!]
\centerline{\includegraphics[width=0.8\textwidth]{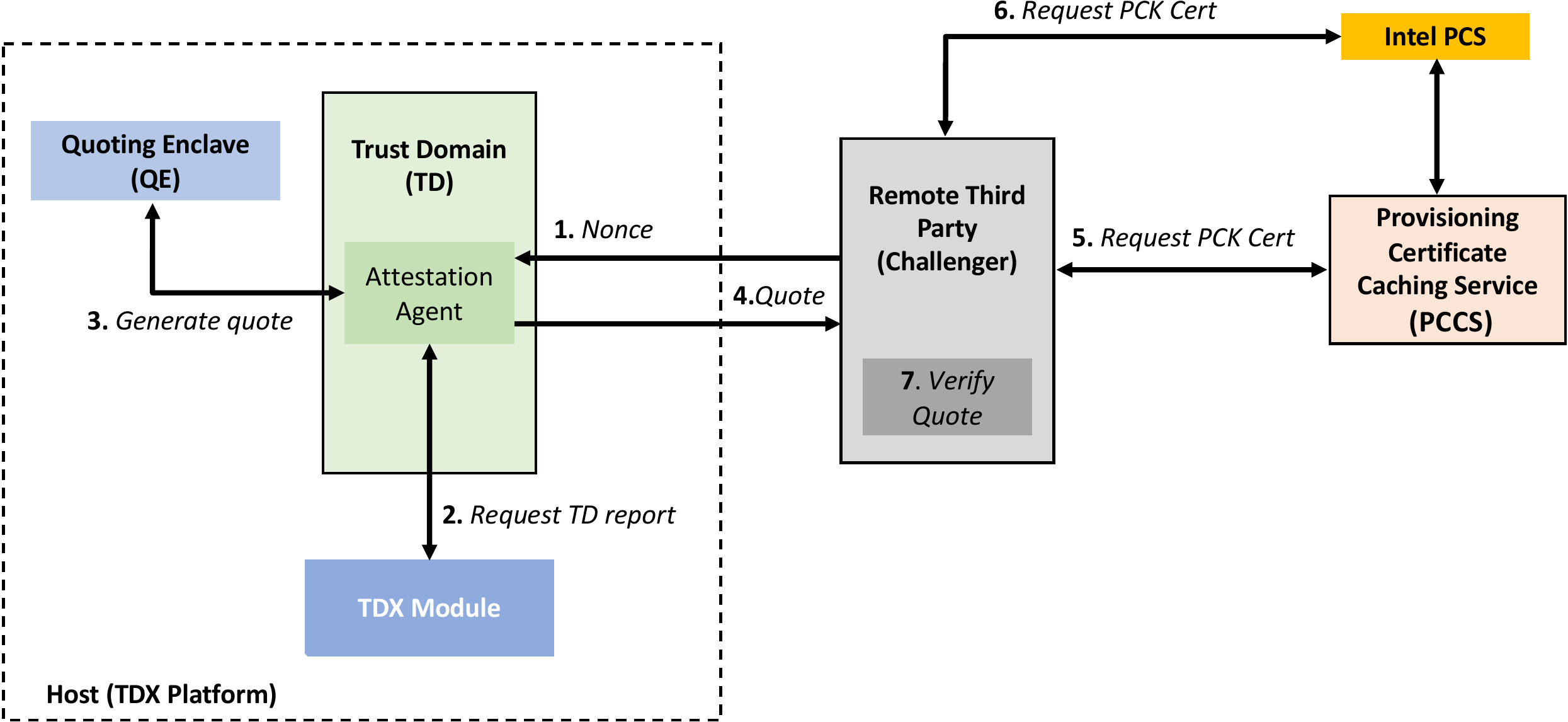}}
\caption{A Remote Attestation Flow}
\label{fig:ra_flow}
\end{figure}
\nip{Remote Attestation Flow.}
\autoref{fig:ra_flow} shows a remote third party performing attestation with an attestation agent running on a \ac{TD}. The remote party sends an attestation request providing a nonce to the attestation agent (Step 1). The nonce provides freshness to the request and prevents replay attacks. The attestation agent retrieves a \ac{TD} \spec{report} from the \ac{TDX} Module providing the nonce as the \code{REPORTDATA} (Step 2) and then subsequently requests the \ac{QE} to sign the \ac{TD} \spec{report} using its attestation key (Step 3).  The \ac{QE} verifies that the \ac{TD} \spec{report} is generated on the platform before signing with its attestation key. The attestation agent then returns the \spec{quote} to the remote party (Step 4). 

The remote party requires the platform's \ac{PCK} certificate to verify the \spec{quote}, so it may download the \ac{PCK} certificate from a \ac{PCCS} (Step 5) or retrieve directly from the Intel \ac{PCS} (Step 6). The party then proceeds to validate the \spec{quote} (Step 7). It checks for the nonce in the \spec{quote} and verifies the integrity of the signature chain from the Intel-issued \ac{PCK} certificate to the signed \spec{quote}, walking the certificate chain to determine whether the \spec{quote} has a valid signature. The party also checks that no keys in the chain have been revoked and whether the \ac{TCB} is up-to-date. Finally, the party checks if the measurements, \ie values in \ac{MRTD} and \acp{RTMR}, in the \spec{quote} match a set of reference values. If it successfully validates the \spec{quote}, the remote party can trust that the \ac{TD} has been properly instantiated on a \ac{TDX} platform. 

\subsection{Use Cases}
\label{ssec:secure_channel}
\nip{Secure Channel Establishment.} 
Remote attestation can be integrated with establishing a secure channel~\cite{knauthsgx}, linking channel setup with the endpoint's TEE identity, state, and configuration. This integration prevents relay
attacks since an attacker cannot forward a challenger's attestation request from a compromised system to a trusted system to service the request.

In a typical scenario when a client negotiates a secure channel with a server running in a \ac{TEE}, it wants to ensure a connection with a properly instantiated server. The server, serving as an attester, generates an ephemeral public and private key pair. It computes the hash of the public key and then creates a \ac{TD} \spec{report} providing this hash as the \code{REPORTDATA}. The server requests a \spec{quote} of the \spec{report} and generates a self-signed certificate with the \spec{quote} embedded in the certificate. It provides this self-signed certificate as the server certificate in the \ac{TLS} handshake protocol. When the client, serving as the challenger, receives the server certificate, it verifies signatures on the certificate and validates the embedded \spec{quote} in the certificate, including the measurements. It also checks if the \spec{quote} includes the hash of the public key since this links the key to the \ac{TEE}. When establishing a secure channel, both client and server can assume the roles of attester and verifier. This allows endpoints running in \acp{TEE} to mutually authenticate each other by validating \acp{TEE}.

\begin{figure}[t!]
\centerline{\includegraphics[width=0.8\textwidth]{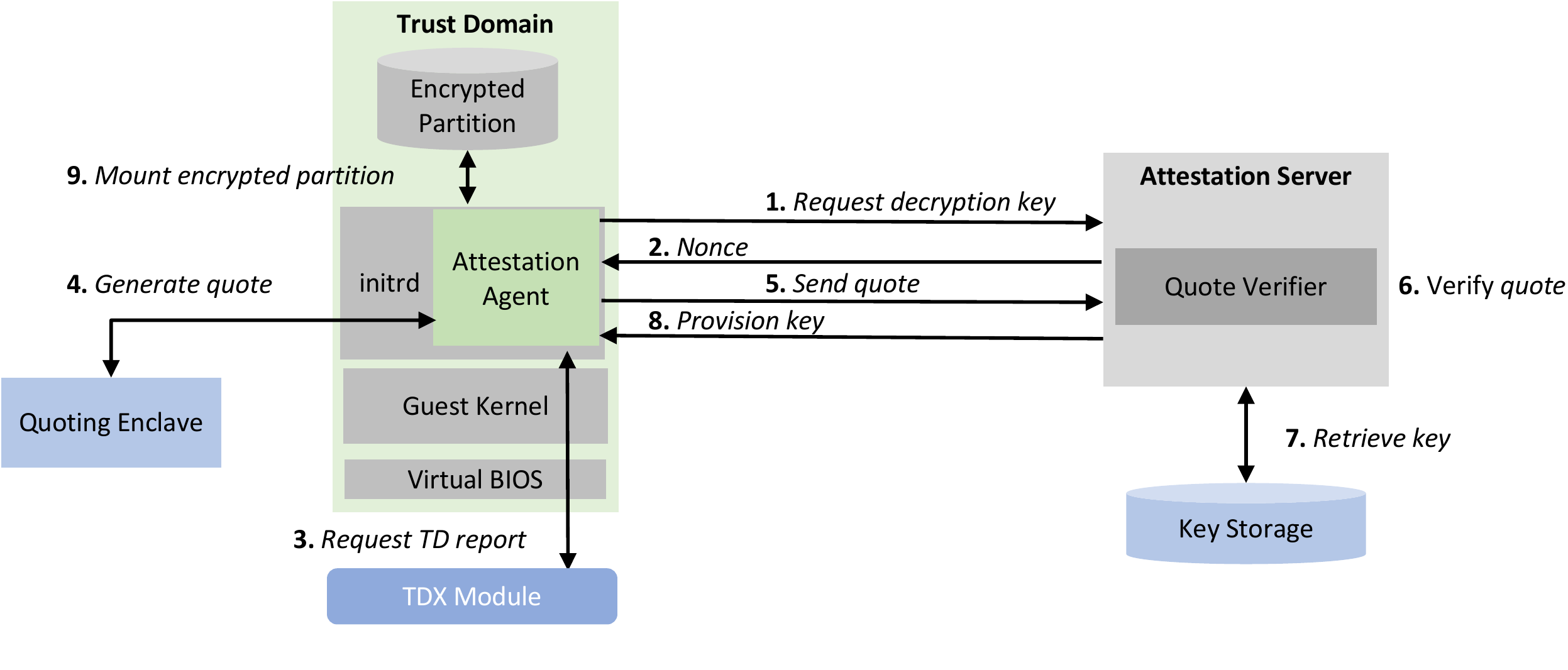}}
\caption{Encrypted Boot Flow}
\label{fig:eb_flow}
\end{figure}
\nip{Encrypted Boot.}
Using remote attestation, we can launch a \ac{TD} with an encrypted partition image and let the tenant control the release of the partition decryption key. \autoref{fig:eb_flow} illustrates the execution flow of an encrypted boot of a \ac{TD}. An attestation agent is placed within the \ac{TD}'s \spec{initrd} and starts when a \ac{TD} is launched. The agent retrieves the \ac{TD}'s \spec{quote}, which contains the \ac{TCB} measurements (from the \ac{TDX} platform up to the \spec{initrd}), and is signed by Intel's \ac{QE}. The \spec{quote} is sent over a secure channel to an attestation server controlled by the tenant for verification. Once a \spec{quote} is verified with the expected measurements, the attestation server provides the decryption key for the partition to the attestation agent. The agent then uses this key to decrypt and mount the encrypted partition that contains the secrets. This ensures that secrets can only be loaded into encrypted memory and are not visible to the host. 

Tenants also have the flexibility to integrate the attestation agent in the virtual BIOS and wrap the entire \ac{VM} workload within an encrypted image. In this case, the virtual BIOS needs to be extended to support the functionalities of retrieving the \spec{quote} and fetching the key for mounting the encrypted disk image.

\section{Conclusion}
\label{sec:conclusion}
In this paper, we provide a top-down review of Intel \ac{TDX}, covering its security principles, threat model, underpinning technologies, system architecture, and future features. We then dive deeper into the design of the \ac{TDX} Module, memory protection mechanisms, and remote attestation. Additionally, we compare \ac{TDX} with similar technologies from other vendors. The review is based on publicly available documentation and source code. As confidential computing is a fast-evolving field, we highlight ongoing challenges and efforts, including the need to support live migration, secure I/O, and extension of trust to peripheral devices and hardware accelerators. We will continue to conduct in-depth security analysis as the technology progresses.
\begin{acks}
We would like to extend our sincere thanks to Guerney Hunt, Rick Boivie, Dimitrios Pendarakis, and Jonathan Bradbury for taking the time to read our draft and providing their invaluable feedback and suggestions.
\end{acks}
\clearpage
\appendix
\section{List of Acronyms}
\label{sec:acronym}
\begin{acronym}[AP-TEE]\itemsep=-3pt
\acro{ABI}{Application Binary Interface}
\acro{ACM}{Authenticated Code Module}
\acro{ACPI}{Advanced Configuration and Power Interface}
\acro{AES}{Advanced Encryption Standard}
\acro{AP-TEE}{Application Platform-Trusted Execution Environment}
\acro{ASID}{Address Space Identifier}
\acro{BIOS}{Basic Input/Output System}
\acro{CCA}{Confidential Compute Architecture}
\acro{CET}{Control Flow Enforcement Technology}
\acro{CMRs}{Convertible Memory Regions}
\acro{CMR}{Convertible Memory Region}
\acro{Ci}{Cryptographic Integrity}
\acro{DCAP}{Data Center Attestation Primitives}
\acro{DHKE}{Diffie-Hellman Key Exchange}
\acro{DMA}{Direct Memory Access}
\acro{DRTM}{Dynamic Root of Trust for Measurement}
\acro{DoS}{Denial of Service}
\acro{ECC}{Error Correction Code}
\acro{EPCM}{Enclave Page Cache Map}
\acro{EPC}{Enclave Page Cache}
\acro{EPT}{Extended Page Table}
\acro{ESM}{Enter Secure Mode}
\acro{ES}{Encrypted State}
\acro{GPA}{Guest Physical Address}
\acro{GPC}{Granule Protection Check}
\acro{GPT}{Granule Protection Table}
\acro{GVA}{Guest Virtual Address}
\acro{HE}{Homomorphic Encryption}
\acro{HKID}{Host Key Identifier}
\acro{HPA}{Host Physical Address}
\acro{IAS}{Intel Attestation Service}
\acro{IOMMU}{Input/Output Memory Management Unit}
\acro{ISA}{Instruction Set Architecture}
\acro{KET}{Key Encryption Table}
\acro{LP}{Logical Processor}
\acro{LRU}{Least Recently Used}
\acro{Li}{Logical Integrity}
\acro{MAC}{Message Authentication Code}
\acro{MEE}{Memory Encryption Engine}
\acro{ME}{Management Engine}
\acro{MKTME}{Multi-key Total Memory Encryption}
\acro{MMIO}{Memory-Mapped Input/Output}
\acro{MMU}{Memory Management Unit}
\acro{MRTD}{Measurement of Trust Domain}
\acro{MSK}{Migration Session Key}
\acro{MSR}{Model-Specific Register}
\acro{MTT}{Memory Tracking Table}
\acro{MigTD}{Migration TD}
\acro{OVMF}{Open Virtual Machine Firmware}
\acro{PAMT}{Physical Address Metadata Table}
\acro{PCCS}{Provisioning Certification Caching service}
\acro{PCE}{Provisioning Certificate Enclave}
\acro{PCI}{Peripheral Component Interconnect}
\acro{PCK}{Provisioning Certification Key}
\acro{PCR}{Platform Configuration Register}
\acro{PCS}{Provisioning Certification Service}
\acro{PEF}{Protected Execution Facility}
\acro{PSP}{Platform Security Processor}
\acro{PTE}{page table entry}
\acro{QE}{Quoting Enclave}
\acro{RME}{Realm Management Extension}
\acro{RMP}{Reverse Mapping Table}
\acro{RTMR}{Runtime Measurement Register}
\acro{S-IOV}{Scalable I/O virtualization}
\acro{SEAM}{Secure-Arbitration Mode}
\acro{SEPT}{Secure EPT}
\acro{SEV}{Secure Encrypted Virtualization}
\acro{SGX}{Software Guard Extensions}
\acro{SLAT}{Second Level Address Translation}
\acro{SMC}{Secure Multi-Party Computation}
\acro{SME}{Secure Memory Encryption}
\acro{SMM}{System Management Mode}
\acro{SNP}{Secure Nested Paging}
\acro{SPR}{Sapphire Rapids}
\acro{SR-IOV}{Single Root I/O virtualization}
\acro{SVM}{Secure Virtual Machine}
\acro{SVN}{Security Version Number}
\acro{SoC}{System-on-Chip}
\acro{TCB}{Trusted Computing Base}
\acro{TDCS}{Trust Domain Control Structure}
\acro{TDMR}{Trust Domain Memory Region}
\acro{TDR}{Trust Domain Root}
\acro{TDVPS}{Trust Domain Virtual Processor State}
\acro{TDX}{Trust Domain Extensions}
\acro{TD}{Trust Domain}
\acro{TEE}{Trusted Execution Environment}
\acro{TLB}{Translation Lookaside Buffer}
\acro{TLS}{Transport Layer Security}
\acro{TME}{Total Memory Encryption}
\acro{TPM}{Trusted Platform Module}
\acro{TSM}{TEE Security Manager}
\acro{TVM}{TEE Virtual Machine}
\acro{TXT}{Trusted Execution Technology}
\acro{UD}{Undefined Instruction}
\acro{UEFI}{Unified Extensible Firmware Interface}
\acro{VE}{Virtualization Exception}
\acro{VMCS}{Virtual Machine Control Structure}
\acro{VMM}{Virtual Machine Monitor}
\acro{VMPL}{Virtual Machine Privilege Level}
\acro{VMX}{Virtual Machine Extensions}
\acro{VM}{Virtual Machine}
\acro{VT}{Virtualization Technology}
\end{acronym}
\clearpage
\bibliographystyle{ACM-Reference-Format}
\bibliography{main}


\begin{thebibliography}{63}


\ifx \showCODEN    \undefined \def \showCODEN     #1{\unskip}     \fi
\ifx \showDOI      \undefined \def \showDOI       #1{#1}\fi
\ifx \showISBNx    \undefined \def \showISBNx     #1{\unskip}     \fi
\ifx \showISBNxiii \undefined \def \showISBNxiii  #1{\unskip}     \fi
\ifx \showISSN     \undefined \def \showISSN      #1{\unskip}     \fi
\ifx \showLCCN     \undefined \def \showLCCN      #1{\unskip}     \fi
\ifx \shownote     \undefined \def \shownote      #1{#1}          \fi
\ifx \showarticletitle \undefined \def \showarticletitle #1{#1}   \fi
\ifx \showURL      \undefined \def \showURL       {\relax}        \fi
\providecommand\bibfield[2]{#2}
\providecommand\bibinfo[2]{#2}
\providecommand\natexlab[1]{#1}
\providecommand\showeprint[2][]{arXiv:#2}

\bibitem[AMD(2020)]%
        {sev2020strengthening}
\bibfield{author}{\bibinfo{person}{AMD}.} \bibinfo{year}{2020}\natexlab{}.
\newblock \showarticletitle{Strengthening VM isolation with integrity
  protection and more}.
\newblock \bibinfo{journal}{\emph{AMD}} (\bibinfo{year}{2020}).
\newblock


\bibitem[Arnautov et~al\mbox{.}(2016)]%
        {arnautov2016scone}
\bibfield{author}{\bibinfo{person}{Sergei Arnautov}, \bibinfo{person}{Bohdan
  Trach}, \bibinfo{person}{Franz Gregor}, \bibinfo{person}{Thomas Knauth},
  \bibinfo{person}{Andre Martin}, \bibinfo{person}{Christian Priebe},
  \bibinfo{person}{Joshua Lind}, \bibinfo{person}{Divya Muthukumaran},
  \bibinfo{person}{Dan O'keeffe}, \bibinfo{person}{Mark Stillwell},
  {et~al\mbox{.}}} \bibinfo{year}{2016}\natexlab{}.
\newblock \showarticletitle{Scone: Secure linux containers with intel sgx}. In
  \bibinfo{booktitle}{\emph{OSDI}}, Vol.~\bibinfo{volume}{16}.
  \bibinfo{pages}{689--703}.
\newblock


\bibitem[Baumann et~al\mbox{.}(2015)]%
        {baumann2015shielding}
\bibfield{author}{\bibinfo{person}{Andrew Baumann}, \bibinfo{person}{Marcus
  Peinado}, {and} \bibinfo{person}{Galen Hunt}.}
  \bibinfo{year}{2015}\natexlab{}.
\newblock \showarticletitle{Shielding applications from an untrusted cloud with
  haven}.
\newblock \bibinfo{journal}{\emph{ACM Transactions on Computer Systems (TOCS)}}
  \bibinfo{volume}{33}, \bibinfo{number}{3} (\bibinfo{year}{2015}),
  \bibinfo{pages}{1--26}.
\newblock


\bibitem[Beekman et~al\mbox{.}(2016)]%
        {beekman2016attestation}
\bibfield{author}{\bibinfo{person}{Jethro~G Beekman}, \bibinfo{person}{John~L
  Manferdelli}, {and} \bibinfo{person}{David Wagner}.}
  \bibinfo{year}{2016}\natexlab{}.
\newblock \showarticletitle{Attestation Transparency: Building secure Internet
  services for legacy clients}. In \bibinfo{booktitle}{\emph{Proceedings of the
  11th ACM on Asia Conference on Computer and Communications Security}}.
  \bibinfo{pages}{687--698}.
\newblock


\bibitem[Behl et~al\mbox{.}(2017)]%
        {behl2017hybrids}
\bibfield{author}{\bibinfo{person}{Johannes Behl}, \bibinfo{person}{Tobias
  Distler}, {and} \bibinfo{person}{R{\"u}diger Kapitza}.}
  \bibinfo{year}{2017}\natexlab{}.
\newblock \showarticletitle{Hybrids on steroids: SGX-based high performance
  BFT}. In \bibinfo{booktitle}{\emph{Proceedings of the Twelfth European
  Conference on Computer Systems}}. \bibinfo{pages}{222--237}.
\newblock


\bibitem[Bohrer et~al\mbox{.}(2004)]%
        {bohrer2004mambo}
\bibfield{author}{\bibinfo{person}{Patrick Bohrer}, \bibinfo{person}{James
  Peterson}, \bibinfo{person}{Mootaz Elnozahy}, \bibinfo{person}{Ram Rajamony},
  \bibinfo{person}{Ahmed Gheith}, \bibinfo{person}{Ron Rockhold},
  \bibinfo{person}{Charles Lefurgy}, \bibinfo{person}{Hazim Shafi},
  \bibinfo{person}{Tarun Nakra}, \bibinfo{person}{Rick Simpson},
  {et~al\mbox{.}}} \bibinfo{year}{2004}\natexlab{}.
\newblock \showarticletitle{Mambo: a full system simulator for the PowerPC
  architecture}.
\newblock \bibinfo{journal}{\emph{ACM SIGMETRICS performance evaluation
  review}} \bibinfo{volume}{31}, \bibinfo{number}{4} (\bibinfo{year}{2004}),
  \bibinfo{pages}{8--12}.
\newblock


\bibitem[Boivie and Williams(2012)]%
        {boivie2012secureblue++}
\bibfield{author}{\bibinfo{person}{Rick Boivie} {and} \bibinfo{person}{Peter
  Williams}.} \bibinfo{year}{2012}\natexlab{}.
\newblock \showarticletitle{SecureBlue++: CPU support for secure execution}.
\newblock \bibinfo{journal}{\emph{IBM, IBM Research Division, RC25287
  (WAT1205-070)}} (\bibinfo{year}{2012}), \bibinfo{pages}{1--9}.
\newblock


\bibitem[Brasser et~al\mbox{.}(2017)]%
        {brasser2017software}
\bibfield{author}{\bibinfo{person}{Ferdinand Brasser}, \bibinfo{person}{Urs
  M{\"u}ller}, \bibinfo{person}{Alexandra Dmitrienko}, \bibinfo{person}{Kari
  Kostiainen}, \bibinfo{person}{Srdjan Capkun}, {and}
  \bibinfo{person}{Ahmad-Reza Sadeghi}.} \bibinfo{year}{2017}\natexlab{}.
\newblock \showarticletitle{Software Grand Exposure: SGX Cache Attacks Are
  Practical.}. In \bibinfo{booktitle}{\emph{WOOT}}. \bibinfo{pages}{11--11}.
\newblock


\bibitem[Brenner et~al\mbox{.}(2016)]%
        {brenner2016securekeeper}
\bibfield{author}{\bibinfo{person}{Stefan Brenner}, \bibinfo{person}{Colin
  Wulf}, \bibinfo{person}{David Goltzsche}, \bibinfo{person}{Nico Weichbrodt},
  \bibinfo{person}{Matthias Lorenz}, \bibinfo{person}{Christof Fetzer},
  \bibinfo{person}{Peter Pietzuch}, {and} \bibinfo{person}{R{\"u}diger
  Kapitza}.} \bibinfo{year}{2016}\natexlab{}.
\newblock \showarticletitle{Securekeeper: confidential zookeeper using intel
  sgx}. In \bibinfo{booktitle}{\emph{Proceedings of the 17th International
  Middleware Conference}}. \bibinfo{pages}{1--13}.
\newblock


\bibitem[Bresticker et~al\mbox{.}(2023)]%
        {riscvcc}
\bibfield{author}{\bibinfo{person}{Andrew Bresticker}, \bibinfo{person}{Andy
  Dellow}, \bibinfo{person}{Atish Patra}, \bibinfo{person}{Atul Khare},
  \bibinfo{person}{Beeman Strong}, \bibinfo{person}{Dingji Li},
  \bibinfo{person}{Dong Du}, \bibinfo{person}{Dylan Reid},
  \bibinfo{person}{Guerney Hunt}, \bibinfo{person}{Kailun Qin},
  \bibinfo{person}{Manuel Offenberg}, \bibinfo{person}{Nick Kossifidis},
  \bibinfo{person}{Ravi Sahita}, \bibinfo{person}{Samuel Ortiz},
  \bibinfo{person}{Vedvyas Shanbhogue}, {and} \bibinfo{person}{Yann Loisel}.}
  \bibinfo{year}{2023}\natexlab{}.
\newblock \showarticletitle{Application-Processor Trusted Execution Environment
  (AP-TEE) for Confidential Computing on RISC-V platforms}.
\newblock  (\bibinfo{year}{2023}).
\newblock


\bibitem[Chen et~al\mbox{.}(2019)]%
        {chen2019sgxpectre}
\bibfield{author}{\bibinfo{person}{Guoxing Chen}, \bibinfo{person}{Sanchuan
  Chen}, \bibinfo{person}{Yuan Xiao}, \bibinfo{person}{Yinqian Zhang},
  \bibinfo{person}{Zhiqiang Lin}, {and} \bibinfo{person}{Ten~H Lai}.}
  \bibinfo{year}{2019}\natexlab{}.
\newblock \showarticletitle{Sgxpectre: Stealing intel secrets from sgx enclaves
  via speculative execution}. In \bibinfo{booktitle}{\emph{2019 IEEE European
  Symposium on Security and Privacy (EuroS\&P)}}. IEEE,
  \bibinfo{pages}{142--157}.
\newblock


\bibitem[Evtyushkin et~al\mbox{.}(2018)]%
        {evtyushkin2018branchscope}
\bibfield{author}{\bibinfo{person}{Dmitry Evtyushkin}, \bibinfo{person}{Ryan
  Riley}, \bibinfo{person}{Nael~CSE Abu-Ghazaleh}, \bibinfo{person}{ECE}, {and}
  \bibinfo{person}{Dmitry Ponomarev}.} \bibinfo{year}{2018}\natexlab{}.
\newblock \showarticletitle{Branchscope: A new side-channel attack on
  directional branch predictor}.
\newblock \bibinfo{journal}{\emph{ACM SIGPLAN Notices}} \bibinfo{volume}{53},
  \bibinfo{number}{2} (\bibinfo{year}{2018}), \bibinfo{pages}{693--707}.
\newblock


\bibitem[Fisch et~al\mbox{.}(2017)]%
        {fisch2017iron}
\bibfield{author}{\bibinfo{person}{Ben Fisch}, \bibinfo{person}{Dhinakaran
  Vinayagamurthy}, \bibinfo{person}{Dan Boneh}, {and} \bibinfo{person}{Sergey
  Gorbunov}.} \bibinfo{year}{2017}\natexlab{}.
\newblock \showarticletitle{Iron: functional encryption using Intel SGX}. In
  \bibinfo{booktitle}{\emph{Proceedings of the 2017 ACM SIGSAC Conference on
  Computer and Communications Security}}. \bibinfo{pages}{765--782}.
\newblock


\bibitem[Fuhry et~al\mbox{.}(2017)]%
        {fuhry2017hardidx}
\bibfield{author}{\bibinfo{person}{Benny Fuhry}, \bibinfo{person}{Raad
  Bahmani}, \bibinfo{person}{Ferdinand Brasser}, \bibinfo{person}{Florian
  Hahn}, \bibinfo{person}{Florian Kerschbaum}, {and}
  \bibinfo{person}{Ahmad-Reza Sadeghi}.} \bibinfo{year}{2017}\natexlab{}.
\newblock \showarticletitle{HardIDX: Practical and secure index with SGX}. In
  \bibinfo{booktitle}{\emph{Data and Applications Security and Privacy XXXI:
  31st Annual IFIP WG 11.3 Conference, DBSec 2017, Philadelphia, PA, USA, July
  19-21, 2017, Proceedings 31}}. Springer, \bibinfo{pages}{386--408}.
\newblock


\bibitem[G{\"o}tzfried et~al\mbox{.}(2017)]%
        {gotzfried2017cache}
\bibfield{author}{\bibinfo{person}{Johannes G{\"o}tzfried},
  \bibinfo{person}{Moritz Eckert}, \bibinfo{person}{Sebastian Schinzel}, {and}
  \bibinfo{person}{Tilo M{\"u}ller}.} \bibinfo{year}{2017}\natexlab{}.
\newblock \showarticletitle{Cache attacks on Intel SGX}. In
  \bibinfo{booktitle}{\emph{Proceedings of the 10th European Workshop on
  Systems Security}}. \bibinfo{pages}{1--6}.
\newblock


\bibitem[Greene(2010)]%
        {txtwhitepaper}
\bibfield{author}{\bibinfo{person}{James Greene}.}
  \bibinfo{year}{2010}\natexlab{}.
\newblock \showarticletitle{{Intel} trusted execution technology:
  {Hardware}-based technology for enhancing server platform security}.
\newblock \bibinfo{journal}{\emph{Intel Corporation}} (\bibinfo{year}{2010}).
\newblock


\bibitem[Gu et~al\mbox{.}(2018)]%
        {gu2018confidential}
\bibfield{author}{\bibinfo{person}{Zhongshu Gu}, \bibinfo{person}{Heqing
  Huang}, \bibinfo{person}{Jialong Zhang}, \bibinfo{person}{Dong Su},
  \bibinfo{person}{Hani Jamjoom}, \bibinfo{person}{Ankita Lamba},
  \bibinfo{person}{Dimitrios Pendarakis}, {and} \bibinfo{person}{Ian Molloy}.}
  \bibinfo{year}{2018}\natexlab{}.
\newblock \showarticletitle{Confidential inference via ternary model
  partitioning}.
\newblock \bibinfo{journal}{\emph{arXiv preprint arXiv:1807.00969}}
  (\bibinfo{year}{2018}).
\newblock


\bibitem[Gu et~al\mbox{.}(2019)]%
        {gu2019reaching}
\bibfield{author}{\bibinfo{person}{Zhongshu Gu}, \bibinfo{person}{Hani
  Jamjoom}, \bibinfo{person}{Dong Su}, \bibinfo{person}{Heqing Huang},
  \bibinfo{person}{Jialong Zhang}, \bibinfo{person}{Tengfei Ma},
  \bibinfo{person}{Dimitrios Pendarakis}, {and} \bibinfo{person}{Ian Molloy}.}
  \bibinfo{year}{2019}\natexlab{}.
\newblock \showarticletitle{Reaching data confidentiality and model
  accountability on the caltrain}. In \bibinfo{booktitle}{\emph{2019 49th
  Annual IEEE/IFIP International Conference on Dependable Systems and Networks
  (DSN)}}. IEEE, \bibinfo{pages}{336--348}.
\newblock


\bibitem[Hunt et~al\mbox{.}(2018)]%
        {pef}
\bibfield{author}{\bibinfo{person}{Guerney Hunt}, \bibinfo{person}{Rick
  Boivie}, \bibinfo{person}{Eric Hall}, \bibinfo{person}{Elaine Palmer},
  \bibinfo{person}{Dimitrios Pendarakis}, {and} \bibinfo{person}{Enriquillo
  Valdez}.} \bibinfo{year}{2018}\natexlab{}.
\newblock \bibinfo{title}{Supporting protected computing on IBM Power
  Architecture}.
\newblock
  \bibinfo{howpublished}{\url{https://developer.ibm.com/articles/l-support-protected-computing/}}.
\newblock


\bibitem[Hunt et~al\mbox{.}(2021)]%
        {hunt2021confidential}
\bibfield{author}{\bibinfo{person}{Guerney~DH Hunt},
  \bibinfo{person}{Ramachandra Pai}, \bibinfo{person}{Michael~V Le},
  \bibinfo{person}{Hani Jamjoom}, \bibinfo{person}{Sukadev Bhattiprolu},
  \bibinfo{person}{Rick Boivie}, \bibinfo{person}{Laurent Dufour},
  \bibinfo{person}{Brad Frey}, \bibinfo{person}{Mohit Kapur},
  \bibinfo{person}{Kenneth~A Goldman}, {et~al\mbox{.}}}
  \bibinfo{year}{2021}\natexlab{}.
\newblock \showarticletitle{Confidential computing for OpenPOWER}. In
  \bibinfo{booktitle}{\emph{Proceedings of the Sixteenth European Conference on
  Computer Systems}}. \bibinfo{pages}{294--310}.
\newblock


\bibitem[IBM(2020)]%
        {ultravisorsrc}
\bibfield{author}{\bibinfo{person}{IBM}.} \bibinfo{year}{2020}\natexlab{}.
\newblock
  \bibinfo{howpublished}{\url{https://github.com/open-power/ultravisor}}.
\newblock


\bibitem[IBM(2022)]%
        {ibmsecureexec}
\bibfield{author}{\bibinfo{person}{IBM}.} \bibinfo{year}{2022}\natexlab{}.
\newblock \showarticletitle{Introducing IBM Secure Execution for Linux 1.3.0}.
\newblock
  \bibinfo{howpublished}{\url{https://www.ibm.com/docs/en/linuxonibm/pdf/l130se03.pdf}}.
\newblock  (\bibinfo{year}{2022}).
\newblock


\bibitem[Intel(2021a)]%
        {cetwhitepaper}
\bibfield{author}{\bibinfo{person}{Intel}.} \bibinfo{year}{2021}\natexlab{a}.
\newblock \showarticletitle{11th Gen Intel® CoreTM vPro® Mobile Platform PCs
  Feature the Industry's Only Silicon-Enabled Threat Protections}.
\newblock  (\bibinfo{year}{2021}).
\newblock


\bibitem[Intel(2021b)]%
        {tdxarchspec}
\bibfield{author}{\bibinfo{person}{Intel}.} \bibinfo{year}{2021}\natexlab{b}.
\newblock \showarticletitle{Intel® Trust Domain CPU Architectural Extensions
  Specification}.
\newblock
  \bibinfo{howpublished}{\url{https://cdrdv2.intel.com/v1/dl/getContent/733582}}.
\newblock  (\bibinfo{year}{2021}).
\newblock


\bibitem[Intel(2021c)]%
        {tdxwhitepaper}
\bibfield{author}{\bibinfo{person}{Intel}.} \bibinfo{year}{2021}\natexlab{c}.
\newblock \showarticletitle{Intel® Trust Domain Extensions}.
\newblock
  \bibinfo{howpublished}{\url{https://cdrdv2.intel.com/v1/dl/getContent/690419}}.
\newblock  (\bibinfo{year}{2021}).
\newblock


\bibitem[Intel(2022a)]%
        {tdxmodulesrc}
\bibfield{author}{\bibinfo{person}{Intel}.} \bibinfo{year}{2022}\natexlab{a}.
\newblock
  \bibinfo{howpublished}{\url{https://www.intel.com/content/www/us/en/download/738875/738876/intel-trust-domain-extension-intel-tdx-module.html}}.
\newblock


\bibitem[Intel(2022b)]%
        {tdxloadersrc}
\bibfield{author}{\bibinfo{person}{Intel}.} \bibinfo{year}{2022}\natexlab{b}.
\newblock
  \bibinfo{howpublished}{\url{https://www.intel.com/content/www/us/en/download/738874/intel-trust-domain-extension-intel-tdx-loader.html}}.
\newblock


\bibitem[Intel(2022c)]%
        {tdxkernelsrc}
\bibfield{author}{\bibinfo{person}{Intel}.} \bibinfo{year}{2022}\natexlab{c}.
\newblock \bibinfo{howpublished}{\url{https://github.com/intel/tdx/}}.
\newblock


\bibitem[Intel(2022d)]%
        {tdxkernelhardening}
\bibfield{author}{\bibinfo{person}{Intel}.} \bibinfo{year}{2022}\natexlab{d}.
\newblock
  \bibinfo{howpublished}{\url{https://intel.github.io/ccc-linux-guest-hardening-docs/security-spec.html}}.
\newblock


\bibitem[Intel(2022e)]%
        {deviceattestation}
\bibfield{author}{\bibinfo{person}{Intel}.} \bibinfo{year}{2022}\natexlab{e}.
\newblock \showarticletitle{Device Attestation Model in Confidential Computing
  Environment}.
\newblock
  \bibinfo{howpublished}{\url{https://cdrdv2.intel.com/v1/dl/getContent/742533}}.
\newblock  (\bibinfo{year}{2022}).
\newblock


\bibitem[Intel(2022f)]%
        {mktmewhitepaper}
\bibfield{author}{\bibinfo{person}{Intel}.} \bibinfo{year}{2022}\natexlab{f}.
\newblock \showarticletitle{Intel® Architecture Memory Encryption
  Technologies}.
\newblock \bibinfo{journal}{\emph{Intel Corporation}} (\bibinfo{year}{2022}).
\newblock


\bibitem[Intel(2022g)]%
        {tdxghci}
\bibfield{author}{\bibinfo{person}{Intel}.} \bibinfo{year}{2022}\natexlab{g}.
\newblock \showarticletitle{Intel® TDX Guest-Hypervisor Communication
  Interface}.
\newblock
  \bibinfo{howpublished}{\url{https://cdrdv2.intel.com/v1/dl/getContent/726790}}.
\newblock  (\bibinfo{year}{2022}).
\newblock


\bibitem[Intel(2022h)]%
        {tdxloaderspec}
\bibfield{author}{\bibinfo{person}{Intel}.} \bibinfo{year}{2022}\natexlab{h}.
\newblock \showarticletitle{Intel® TDX Loader Interface Specification}.
\newblock
  \bibinfo{howpublished}{\url{https://cdrdv2.intel.com/v1/dl/getContent/733584}}.
\newblock  (\bibinfo{year}{2022}).
\newblock


\bibitem[Intel(2022i)]%
        {tdxmodulespec}
\bibfield{author}{\bibinfo{person}{Intel}.} \bibinfo{year}{2022}\natexlab{i}.
\newblock \showarticletitle{Intel® TDX Module 1.0 Specification}.
\newblock
  \bibinfo{howpublished}{\url{https://cdrdv2.intel.com/v1/dl/getContent/733568}}.
\newblock  (\bibinfo{year}{2022}).
\newblock


\bibitem[Intel(2022j)]%
        {tdxvirtualfw}
\bibfield{author}{\bibinfo{person}{Intel}.} \bibinfo{year}{2022}\natexlab{j}.
\newblock \showarticletitle{Intel® TDX Virtual Firmware Design Guide}.
\newblock
  \bibinfo{howpublished}{\url{https://cdrdv2.intel.com/v1/dl/getContent/733585}}.
\newblock  (\bibinfo{year}{2022}).
\newblock


\bibitem[Intel(2022k)]%
        {tdxio}
\bibfield{author}{\bibinfo{person}{Intel}.} \bibinfo{year}{2022}\natexlab{k}.
\newblock \showarticletitle{Software Enabling for Intel® TDX in Support of
  TEE-I/O}.
\newblock
  \bibinfo{howpublished}{\url{https://cdrdv2.intel.com/v1/dl/getContent/742542}}.
\newblock  (\bibinfo{year}{2022}).
\newblock


\bibitem[Intel(2022l)]%
        {tdvf}
\bibfield{author}{\bibinfo{person}{Intel}.} \bibinfo{year}{2022}\natexlab{l}.
\newblock \bibinfo{title}{TDX Virtual Firmware (TDVF)}.
\newblock
  \bibinfo{howpublished}{\url{https://github.com/tianocore/edk2-staging/tree/TDVF}}.
\newblock


\bibitem[Intel(2023a)]%
        {tdx-migration}
\bibfield{author}{\bibinfo{person}{Intel}.} \bibinfo{year}{2023}\natexlab{a}.
\newblock \bibinfo{title}{Intel® TDX Module Architecture Specification: TD
  Migration}.
\newblock
  \bibinfo{howpublished}{\url{https://cdrdv2.intel.com/v1/dl/getContent/733578}}.
\newblock


\bibitem[Intel(2023b)]%
        {tdpartitioning}
\bibfield{author}{\bibinfo{person}{Intel}.} \bibinfo{year}{2023}\natexlab{b}.
\newblock \showarticletitle{Intel® Trust Domain Extensions (Intel® TDX)
  Module TD Partitioning Architecture Specification}.
\newblock
  \bibinfo{howpublished}{\url{https://cdrdv2.intel.com/v1/dl/getContent/773039}}.
\newblock  (\bibinfo{year}{2023}).
\newblock


\bibitem[Kaplan(2017)]%
        {kaplan2017protecting}
\bibfield{author}{\bibinfo{person}{David Kaplan}.}
  \bibinfo{year}{2017}\natexlab{}.
\newblock \showarticletitle{Protecting vm register state with sev-es}.
\newblock \bibinfo{journal}{\emph{AMD}} (\bibinfo{year}{2017}).
\newblock


\bibitem[Kaplan et~al\mbox{.}(2016)]%
        {kaplan2016amd}
\bibfield{author}{\bibinfo{person}{David Kaplan}, \bibinfo{person}{Jeremy
  Powell}, {and} \bibinfo{person}{Tom Woller}.}
  \bibinfo{year}{2016}\natexlab{}.
\newblock \showarticletitle{AMD memory encryption}.
\newblock \bibinfo{journal}{\emph{AMD}} (\bibinfo{year}{2016}).
\newblock


\bibitem[Kim et~al\mbox{.}(2014)]%
        {rowhammer-ieee-2014}
\bibfield{author}{\bibinfo{person}{Yoongu Kim}, \bibinfo{person}{Ross Daly},
  \bibinfo{person}{Jeremie Kim}, \bibinfo{person}{Chris Fallin},
  \bibinfo{person}{Ji~Hye Lee}, \bibinfo{person}{Donghyuk Lee},
  \bibinfo{person}{Chris Wilkerson}, \bibinfo{person}{Konrad Lai}, {and}
  \bibinfo{person}{Onur Mutlu}.} \bibinfo{year}{2014}\natexlab{}.
\newblock \showarticletitle{Flipping bits in memory without accessing them: An
  experimental study of DRAM disturbance errors}. In
  \bibinfo{booktitle}{\emph{2014 ACM/IEEE 41st International Symposium on
  Computer Architecture (ISCA)}}. \bibinfo{pages}{361--372}.
\newblock
\urldef\tempurl%
\url{https://doi.org/10.1109/ISCA.2014.6853210}
\showDOI{\tempurl}


\bibitem[Knauth et~al\mbox{.}(2019)]%
        {knauthsgx}
\bibfield{author}{\bibinfo{person}{Thomas Knauth}, \bibinfo{person}{Michael
  Steiner}, \bibinfo{person}{Somnath Chakrabarti}, \bibinfo{person}{Li Lei},
  \bibinfo{person}{Cedric Xing}, {and} \bibinfo{person}{Mona Vij}.}
  \bibinfo{year}{2019}\natexlab{}.
\newblock \showarticletitle{Integrating Intel SGX Remote Attestation with
  Transport Layer Security}.
\newblock \bibinfo{journal}{\emph{arXiv preprint arXiv:1801.05863v2}}
  (\bibinfo{year}{2019}).
\newblock


\bibitem[Koruyeh et~al\mbox{.}(2018)]%
        {koruyeh2018spectre}
\bibfield{author}{\bibinfo{person}{Esmaeil~Mohammadian Koruyeh},
  \bibinfo{person}{Khaled Khasawneh}, \bibinfo{person}{Chengyu Song}, {and}
  \bibinfo{person}{Nael Abu-Ghazaleh}.} \bibinfo{year}{2018}\natexlab{}.
\newblock \showarticletitle{Spectre Returns! Speculation Attacks using the
  Return Stack Buffer}. In \bibinfo{booktitle}{\emph{12th USENIX Workshop on
  Offensive Technologies}}.
\newblock


\bibitem[Lee et~al\mbox{.}(2017)]%
        {lee2017inferring}
\bibfield{author}{\bibinfo{person}{Sangho Lee}, \bibinfo{person}{Ming-Wei
  Shih}, \bibinfo{person}{Prasun Gera}, \bibinfo{person}{Taesoo Kim},
  \bibinfo{person}{Hyesoon Kim}, {and} \bibinfo{person}{Marcus Peinado}.}
  \bibinfo{year}{2017}\natexlab{}.
\newblock \showarticletitle{Inferring Fine-grained Control Flow Inside SGX
  Enclaves with Branch Shadowing.}. In \bibinfo{booktitle}{\emph{USENIX
  Security Symposium}}, Vol.~\bibinfo{volume}{19}. \bibinfo{pages}{16--18}.
\newblock


\bibitem[Li et~al\mbox{.}(2022)]%
        {li2022design}
\bibfield{author}{\bibinfo{person}{Xupeng Li}, \bibinfo{person}{Xuheng Li},
  \bibinfo{person}{Christoffer Dall}, \bibinfo{person}{Ronghui Gu},
  \bibinfo{person}{Jason Nieh}, \bibinfo{person}{Yousuf Sait}, {and}
  \bibinfo{person}{Gareth Stockwell}.} \bibinfo{year}{2022}\natexlab{}.
\newblock \showarticletitle{Design and Verification of the Arm Confidential
  Compute Architecture}. In \bibinfo{booktitle}{\emph{16th USENIX Symposium on
  Operating Systems Design and Implementation (OSDI 22)}}.
  \bibinfo{pages}{465--484}.
\newblock


\bibitem[Lind et~al\mbox{.}(2017)]%
        {lind2017glamdring}
\bibfield{author}{\bibinfo{person}{Joshua Lind}, \bibinfo{person}{Christian
  Priebe}, \bibinfo{person}{Divya Muthukumaran}, \bibinfo{person}{Dan
  O{\textquoteright}Keeffe}, \bibinfo{person}{Pierre-Louis Aublin},
  \bibinfo{person}{Florian Kelbert}, \bibinfo{person}{Tobias Reiher},
  \bibinfo{person}{David Goltzsche}, \bibinfo{person}{David Eyers},
  \bibinfo{person}{R{\"u}diger Kapitza}, \bibinfo{person}{Christof Fetzer},
  {and} \bibinfo{person}{Peter Pietzuch}.} \bibinfo{year}{2017}\natexlab{}.
\newblock \showarticletitle{Glamdring: Automatic Application Partitioning for
  Intel {SGX}}. In \bibinfo{booktitle}{\emph{2017 USENIX Annual Technical
  Conference (USENIX ATC 17)}}. \bibinfo{publisher}{USENIX Association},
  \bibinfo{address}{Santa Clara, CA}, \bibinfo{pages}{285--298}.
\newblock


\bibitem[McKeen et~al\mbox{.}(2013)]%
        {mckeen2013innovative}
\bibfield{author}{\bibinfo{person}{Frank McKeen}, \bibinfo{person}{Ilya
  Alexandrovich}, \bibinfo{person}{Alex Berenzon}, \bibinfo{person}{Carlos~V
  Rozas}, \bibinfo{person}{Hisham Shafi}, \bibinfo{person}{Vedvyas Shanbhogue},
  {and} \bibinfo{person}{Uday~R Savagaonkar}.} \bibinfo{year}{2013}\natexlab{}.
\newblock \showarticletitle{Innovative instructions and software model for
  isolated execution.}
\newblock \bibinfo{journal}{\emph{Hasp@ isca}} \bibinfo{volume}{10},
  \bibinfo{number}{1} (\bibinfo{year}{2013}).
\newblock


\bibitem[Moghimi et~al\mbox{.}(2017)]%
        {moghimi2017cachezoom}
\bibfield{author}{\bibinfo{person}{Ahmad Moghimi}, \bibinfo{person}{Gorka
  Irazoqui}, {and} \bibinfo{person}{Thomas Eisenbarth}.}
  \bibinfo{year}{2017}\natexlab{}.
\newblock \showarticletitle{Cachezoom: How SGX amplifies the power of cache
  attacks}. In \bibinfo{booktitle}{\emph{Cryptographic Hardware and Embedded
  Systems--CHES 2017: 19th International Conference, Taipei, Taiwan, September
  25-28, 2017, Proceedings}}. Springer, \bibinfo{pages}{69--90}.
\newblock


\bibitem[Ohrimenko et~al\mbox{.}(2016)]%
        {ohrimenko2016oblivious}
\bibfield{author}{\bibinfo{person}{Olga Ohrimenko}, \bibinfo{person}{Felix
  Schuster}, \bibinfo{person}{C{\'e}dric Fournet}, \bibinfo{person}{Aastha
  Mehta}, \bibinfo{person}{Sebastian Nowozin}, \bibinfo{person}{Kapil Vaswani},
  {and} \bibinfo{person}{Manuel Costa}.} \bibinfo{year}{2016}\natexlab{}.
\newblock \showarticletitle{Oblivious multi-party machine learning on trusted
  processors.}. In \bibinfo{booktitle}{\emph{USENIX Security Symposium}},
  Vol.~\bibinfo{volume}{16}. \bibinfo{pages}{10--12}.
\newblock


\bibitem[Priebe et~al\mbox{.}(2018)]%
        {priebe2018enclavedb}
\bibfield{author}{\bibinfo{person}{Christian Priebe}, \bibinfo{person}{Kapil
  Vaswani}, {and} \bibinfo{person}{Manuel Costa}.}
  \bibinfo{year}{2018}\natexlab{}.
\newblock \showarticletitle{EnclaveDB: A secure database using SGX}. In
  \bibinfo{booktitle}{\emph{2018 IEEE Symposium on Security and Privacy (SP)}}.
  IEEE, \bibinfo{pages}{264--278}.
\newblock


\bibitem[Scarlata et~al\mbox{.}(2018)]%
        {dcapwhitepaper}
\bibfield{author}{\bibinfo{person}{Vinnie Scarlata}, \bibinfo{person}{Simon
  Johnson}, \bibinfo{person}{James Beaney}, {and} \bibinfo{person}{Piotr
  Zmijewski}.} \bibinfo{year}{2018}\natexlab{}.
\newblock \showarticletitle{Supporting third party attestation for Intel SGX
  with Intel data center attestation primitives}.
\newblock
  \bibinfo{howpublished}{\url{https://cdrdv2-public.intel.com/671314/intel-sgx-support-for-third-party-attestation.pdf}}.
\newblock  (\bibinfo{year}{2018}).
\newblock


\bibitem[Schuster et~al\mbox{.}(2015)]%
        {schuster2015vc3}
\bibfield{author}{\bibinfo{person}{Felix Schuster}, \bibinfo{person}{Manuel
  Costa}, \bibinfo{person}{C{\'e}dric Fournet}, \bibinfo{person}{Christos
  Gkantsidis}, \bibinfo{person}{Marcus Peinado}, \bibinfo{person}{Gloria
  Mainar-Ruiz}, {and} \bibinfo{person}{Mark Russinovich}.}
  \bibinfo{year}{2015}\natexlab{}.
\newblock \showarticletitle{VC3: Trustworthy data analytics in the cloud using
  SGX}. In \bibinfo{booktitle}{\emph{2015 IEEE symposium on security and
  privacy}}. IEEE, \bibinfo{pages}{38--54}.
\newblock


\bibitem[Schwarz et~al\mbox{.}(2017)]%
        {schwarz2017malware}
\bibfield{author}{\bibinfo{person}{Michael Schwarz}, \bibinfo{person}{Samuel
  Weiser}, \bibinfo{person}{Daniel Gruss}, \bibinfo{person}{Cl{\'e}mentine
  Maurice}, {and} \bibinfo{person}{Stefan Mangard}.}
  \bibinfo{year}{2017}\natexlab{}.
\newblock \showarticletitle{Malware guard extension: Using SGX to conceal cache
  attacks}. In \bibinfo{booktitle}{\emph{Detection of Intrusions and Malware,
  and Vulnerability Assessment: 14th International Conference, DIMVA 2017,
  Bonn, Germany, July 6-7, 2017, Proceedings 14}}. Springer,
  \bibinfo{pages}{3--24}.
\newblock


\bibitem[Thomas~Lendacky(2023)]%
        {linuxsvsm}
\bibfield{author}{\bibinfo{person}{Carlos~Bilbao Thomas~Lendacky}.}
  \bibinfo{year}{2023}\natexlab{}.
\newblock \bibinfo{title}{Linux SVSM (Secure VM Service Module)}.
\newblock \bibinfo{howpublished}{\url{https://github.com/AMDESE/linux-svsm}}.
\newblock


\bibitem[Tramer and Boneh(2018)]%
        {tramer2018slalom}
\bibfield{author}{\bibinfo{person}{Florian Tramer} {and} \bibinfo{person}{Dan
  Boneh}.} \bibinfo{year}{2018}\natexlab{}.
\newblock \showarticletitle{Slalom: Fast, verifiable and private execution of
  neural networks in trusted hardware}.
\newblock \bibinfo{journal}{\emph{arXiv preprint arXiv:1806.03287}}
  (\bibinfo{year}{2018}).
\newblock


\bibitem[Tsai et~al\mbox{.}(2017)]%
        {tsai2017graphene}
\bibfield{author}{\bibinfo{person}{Chia-Che Tsai}, \bibinfo{person}{Donald~E
  Porter}, {and} \bibinfo{person}{Mona Vij}.} \bibinfo{year}{2017}\natexlab{}.
\newblock \showarticletitle{Graphene-SGX: A Practical Library OS for Unmodified
  Applications on SGX.}. In \bibinfo{booktitle}{\emph{USENIX Annual Technical
  Conference}}. \bibinfo{pages}{645--658}.
\newblock


\bibitem[Uhlig et~al\mbox{.}(2005)]%
        {vtx}
\bibfield{author}{\bibinfo{person}{Rich Uhlig}, \bibinfo{person}{Gil Neiger},
  \bibinfo{person}{Dion Rodgers}, \bibinfo{person}{Amy~L Santoni},
  \bibinfo{person}{Fernando~CM Martins}, \bibinfo{person}{Andrew~V Anderson},
  \bibinfo{person}{Steven~M Bennett}, \bibinfo{person}{Alain Kagi},
  \bibinfo{person}{Felix~H Leung}, {and} \bibinfo{person}{Larry Smith}.}
  \bibinfo{year}{2005}\natexlab{}.
\newblock \showarticletitle{Intel virtualization technology}.
\newblock \bibinfo{journal}{\emph{Computer}} (\bibinfo{year}{2005}).
\newblock


\bibitem[Van~Bulck et~al\mbox{.}(2017a)]%
        {van2017sgx}
\bibfield{author}{\bibinfo{person}{Jo Van~Bulck}, \bibinfo{person}{Frank
  Piessens}, {and} \bibinfo{person}{Raoul Strackx}.}
  \bibinfo{year}{2017}\natexlab{a}.
\newblock \showarticletitle{SGX-Step: A practical attack framework for precise
  enclave execution control}. In \bibinfo{booktitle}{\emph{Proceedings of the
  2nd Workshop on System Software for Trusted Execution}}.
  \bibinfo{pages}{1--6}.
\newblock


\bibitem[Van~Bulck et~al\mbox{.}(2017b)]%
        {van2017telling}
\bibfield{author}{\bibinfo{person}{Jo Van~Bulck}, \bibinfo{person}{Nico
  Weichbrodt}, \bibinfo{person}{R{\"u}diger Kapitza}, \bibinfo{person}{Frank
  Piessens}, {and} \bibinfo{person}{Raoul Strackx}.}
  \bibinfo{year}{2017}\natexlab{b}.
\newblock \showarticletitle{Telling your secrets without page faults: Stealthy
  page table-based attacks on enclaved execution}. In
  \bibinfo{booktitle}{\emph{Proceedings of the 26th USENIX Security
  Symposium}}. USENIX Association, \bibinfo{pages}{1041--1056}.
\newblock


\bibitem[Wang et~al\mbox{.}(2017)]%
        {wang2017leaky}
\bibfield{author}{\bibinfo{person}{Wenhao Wang}, \bibinfo{person}{Guoxing
  Chen}, \bibinfo{person}{Xiaorui Pan}, \bibinfo{person}{Yinqian Zhang},
  \bibinfo{person}{XiaoFeng Wang}, \bibinfo{person}{Vincent Bindschaedler},
  \bibinfo{person}{Haixu Tang}, {and} \bibinfo{person}{Carl~A Gunter}.}
  \bibinfo{year}{2017}\natexlab{}.
\newblock \showarticletitle{Leaky cauldron on the dark land: Understanding
  memory side-channel hazards in SGX}. In \bibinfo{booktitle}{\emph{Proceedings
  of the 2017 ACM SIGSAC Conference on Computer and Communications Security}}.
  \bibinfo{pages}{2421--2434}.
\newblock


\bibitem[Williams and Boivie(2011)]%
        {williams2011cpu}
\bibfield{author}{\bibinfo{person}{Peter Williams} {and} \bibinfo{person}{Rick
  Boivie}.} \bibinfo{year}{2011}\natexlab{}.
\newblock \showarticletitle{CPU support for secure executables}. In
  \bibinfo{booktitle}{\emph{Trust and Trustworthy Computing: 4th International
  Conference, TRUST 2011, Pittsburgh, PA, USA, June 22-24, 2011. Proceedings
  4}}. Springer, \bibinfo{pages}{172--187}.
\newblock


\bibitem[Xu et~al\mbox{.}(2015)]%
        {xu2015controlled}
\bibfield{author}{\bibinfo{person}{Yuanzhong Xu}, \bibinfo{person}{Weidong
  Cui}, {and} \bibinfo{person}{Marcus Peinado}.}
  \bibinfo{year}{2015}\natexlab{}.
\newblock \showarticletitle{Controlled-channel attacks: Deterministic side
  channels for untrusted operating systems}. In \bibinfo{booktitle}{\emph{2015
  IEEE Symposium on Security and Privacy}}. IEEE, \bibinfo{pages}{640--656}.
\newblock


\end{thebibliography}
\end{document}